\documentclass[
superscriptaddress,aps,preprintnumbers,amsmath,amssymb,prd,nofootinbib,preprint]{revtex4-1}

\usepackage{graphicx}
\usepackage{epstopdf}
\usepackage{dcolumn}
\usepackage{bm}
\usepackage{hyperref}
\usepackage{color}
\usepackage{amsmath}

\begin{document}


\def\a{\alpha}
\def\b{\beta}
\def\c{\varepsilon}
\def\d{\delta}
\def\e{\epsilon}
\def\f{\phi}
\def\g{\gamma}
\def\h{\theta}
\def\k{\kappa}
\def\l{\lambda}
\def\m{\mu}
\def\n{\nu}
\def\p{\psi}
\def\q{\partial}
\def\r{\rho}
\def\s{\sigma}
\def\t{\tau}
\def\u{\upsilon}
\def\v{\varphi}
\def\w{\omega}
\def\x{\xi}
\def\y{\eta}
\def\z{\zeta}
\def\D{\Delta}
\def\G{\Gamma}
\def\H{\Theta}
\def\L{\Lambda}
\def\F{\Phi}
\def\P{\Psi}
\def\S{\Sigma}

\def\o{\over}
\def\beq{\begin{align}}
\def\eeq{\end{align}}
\newcommand{\gsim}{ \mathop{}_{\textstyle \sim}^{\textstyle >} }
\newcommand{\lsim}{ \mathop{}_{\textstyle \sim}^{\textstyle <} }
\newcommand{\vev}[1]{ \left\langle {#1} \right\rangle }
\newcommand{\bra}[1]{ \langle {#1} | }
\newcommand{\ket}[1]{ | {#1} \rangle }
\newcommand{\EV}{ {\rm eV} }
\newcommand{\KEV}{ {\rm keV} }
\newcommand{\MEV}{ {\rm MeV} }
\newcommand{\GEV}{ {\rm GeV} }
\newcommand{\TEV}{ {\rm TeV} }
\newcommand{\1}{\mbox{1}\hspace{-0.25em}\mbox{l}}
\newcommand{\headline}[1]{\noindent{\bf #1}}
\def\diag{\mathop{\rm diag}\nolimits}
\def\Spin{\mathop{\rm Spin}}
\def\SO{\mathop{\rm SO}}
\def\O{\mathop{\rm O}}
\def\SU{\mathop{\rm SU}}
\def\U{\mathop{\rm U}}
\def\Sp{\mathop{\rm Sp}}
\def\SL{\mathop{\rm SL}}
\def\tr{\mathop{\rm tr}}
\def\mpl{M_{\rm Pl}}

\def\IJMP{Int.~J.~Mod.~Phys. }
\def\MPL{Mod.~Phys.~Lett. }
\def\NP{Nucl.~Phys. }
\def\PL{Phys.~Lett. }
\def\PR{Phys.~Rev. }
\def\PRL{Phys.~Rev.~Lett. }
\def\PTP{Prog.~Theor.~Phys. }
\def\ZP{Z.~Phys. }

\def\dd{\mathrm{d}}
\def\ff{\mathrm{f}}
\def\BH{{\rm BH}}
\def\inf{{\rm inf}}
\def\ev{{\rm evap}}
\def\eq{{\rm eq}}
\def\SM{{\rm sm}}
\def\Mpl{M_{\rm Pl}}
\def\GeV{{\rm GeV}}
\newcommand{\Red}[1]{\textcolor{red}{#1}}
\newcommand{\TL}[1]{\textcolor{blue}{\bf TL: #1}}


\hfill SCIPP 17/04 \\

\title{
Saxion Cosmology for Thermalized Gravitino Dark Matter
}

\author{Raymond T. Co}
\affiliation{Department of Physics, University of California, Berkeley, California 94720, USA}
\affiliation{Theoretical Physics Group, Lawrence Berkeley National Laboratory, Berkeley, California 94720, USA}
\author{Francesco D'Eramo}
\affiliation{Department of Physics, University of California Santa Cruz, Santa Cruz, CA 95064, USA}
\affiliation{Santa Cruz Institute for Particle Physics, Santa Cruz, CA 95064, USA}
\author{Lawrence J. Hall}
\affiliation{Department of Physics, University of California, Berkeley, California 94720, USA}
\affiliation{Theoretical Physics Group, Lawrence Berkeley National Laboratory, Berkeley, California 94720, USA}
\author{Keisuke Harigaya}
\affiliation{Department of Physics, University of California, Berkeley, California 94720, USA}
\affiliation{Theoretical Physics Group, Lawrence Berkeley National Laboratory, Berkeley, California 94720, USA}

\begin{abstract}
In all supersymmetric theories, gravitinos, with mass suppressed by the Planck scale, are an obvious candidate for dark matter; but if gravitinos ever reached thermal equilibrium, such dark matter is apparently either too abundant or too hot, and is excluded.   However, in theories with an axion, a saxion condensate is generated during an early era of cosmological history and its late decay dilutes dark matter.  We show that such dilution allows previously thermalized gravitinos to account for the observed dark matter over very wide ranges of gravitino mass, keV $ < m_{3/2} < $ TeV, axion decay constant, $10^9 \, \GeV < f_a < 10^{16} \, \GeV$, and saxion mass, 10 MeV $< m_s < $ 100 TeV.  Constraints on this parameter space are studied from BBN, supersymmetry breaking, gravitino and axino production from freeze-in and saxion decay, and from axion production from both misalignment and parametric resonance mechanisms.  Large allowed regions of $(m_{3/2}, f_a, m_s)$ remain, but differ for DFSZ and KSVZ theories.  Superpartner production at colliders may lead to events with displaced vertices and kinks, and may contain saxions decaying to $(WW,ZZ,hh) ,gg, \gamma \gamma$ or a pair of Standard Model fermions.  Freeze-in may lead to a sub-dominant warm component of gravitino dark matter, and saxion decay to axions may lead to dark radiation. 
\end{abstract}

\date{\today}

\maketitle
 
\tableofcontents

\newpage

\section{Introduction}

If supersymmetry is relevant for the hierarchy problem, gravitinos, with a mass suppressed by the Planck mass, become an interesting candidate for dark matter, as pointed out by Witten~\cite{Witten:1981nf}.  However, the cosmology of gravitinos has long been viewed as problematic.
In 1981 Pagels and Primack found that light gravitinos would overclose the universe if they were heavier than the keV scale \cite{Pagels:1981ke}.
To obtain the dark matter abundance revealed by recent measurements, the gravitino mass must be around 100 eV, which is excluded due to the warmness of the gravitino~\cite{Viel:2005qj}.
These pioneering works assumed that gravitinos, like photons and neutrinos, would be in thermal equilibrium in the very early universe with a high temperature, so that their number density would be given by thermodynamics.  Since then it has typically been assumed that, in theories with weak scale supersymmetry, the reheat temperature of the universe after inflation $T_{\rm R}$ is severely restricted \cite{Moroi:1993mb}, to strongly limit the gravitino abundance.   

However, in supersymmetric theories with a Peccei-Quinn (PQ) symmetry~\cite{Peccei:1977hh} broken at scale $V_{\rm PQ}$ to solve the strong CP problem, the gravitino abundance can be diluted by the late decay of a saxion condensate~\cite{Banks:2002sd,Kawasaki:2008jc} which is generated by supersymmetry breaking during an early era, for example during inflation \cite{Dine:1995uk}. (See~\cite{Kim:1992eu,Lyth:1993zw} for dilution by thermally produced saxions.)  Hence, in this paper we return to the original assumption of Witten, Pagels and Primack that gravitinos were in thermal equilibrium in the very early universe.  We take the gravitino to be the Lightest Supersymmetric Partner (LSP) and study constraints on such gravitino dark matter, exploring which regions of the $(m_{3/2}, V_{\rm PQ})$ plane are preferred and hence relating the scales of supersymmetry and PQ breaking.
Most constraints are independent of the very early cosmological history of the saxion oscillations, depending only on the saxion evolution at temperatures less than or of order of the masses of superpartners, especially those of the higgsino and lightest observable supersymmetric particle (LOSP).

Dilution from the saxion condensate allows a much wider set of cosmologies, in particular allowing $T_ {\rm R}$ to be arbitrarily high.  For comparison, without dilution a light gravitino, $m_{3/2} < $~MeV, requires $T_{\rm R}$ below the TeV scale of superpartners.  Thus saxion dilution allows high $T_{\rm R}$ scenaria for the interesting case of displaced vertex signals at LHC, and its MATHUSLA extension~\cite{Chou:2016lxi}, arising from decays to gravitinos.

Cosmological axion production from the misalignment mechanism~\cite{Preskill:1982cy,Abbott:1982af,Dine:1982ah} is frequently taken to limit $V_{\rm PQ}=f_a N_{\rm DW}/\sqrt{2} \lsim  N_{\rm DW} \times 10^{12}$ GeV, where $f_a$ is the axion decay constant and $N_{\rm DW}$ is the domain wall number.  However, with dilution from a saxion condensate this limit is  weakened by 3-4 orders of magnitude \cite{Lazarides:1987zf,Kawasaki:1995vt,Hashimoto:1998ua}.  Hence we also explore the abundance of axion dark matter, finding regions of parameter space in both DFSZ \cite{Dine:1981rt, Zhitnitsky:1980tq} and KSVZ \cite{Kim:1979if, Shifman:1979if} models where it can be a significant component of dark matter.

\section{The Cosmological History}
\label{sec:history}

In this section we provide an overview of the cosmological evolution of the saxion condensate and the thermal bath, and we give results for the axion abundance.

In the absence of supersymmetry breaking, the saxion field $s$ has no potential. In the early universe, at any era the non-zero energy density of the universe breaks supersymmetry, and hence the form of the saxion potential is a highly model-dependent question.   As the universe evolves through inflation, post-inflation, reheating and subsequent eras the saxion potential and its minimum changes leading, in general, to a highly complicated evolution of the saxion condensate.  Rather than studying a particular model, we show that the physics relevant for gravitino dark matter depends on the saxion evolution only at temperatures less than or of order $\tilde{m}$, the masses of the SM superpartners, specifically the masses of the higgsino and the LOSP, which we take to be $\cal{O}$(TeV).  Thus, to obtain the main results of this paper the assumption on the cosmological history of the universe at temperatures above the TeV scale is extremely mild:
\begin{itemize}
\item Before reaching the TeV scale, the saxion field acquired a large displacement from its present value and there was an era where gravitinos were in thermal equilibrium.
\end{itemize}
Furthermore, for later evolution of the saxion field we assume
\begin{itemize}
\item  From $T \sim \tilde{m} \sim$ TeV until it decays, the saxion condensate oscillates about the present vacuum in a quadratic potential 
\begin{align}
V =\frac{1}{2} \; m_s^2 s^2
\label{eq:quadsaxpot}
\end{align}
where $m_s$ is the soft supersymmetry breaking mass of the saxion.
\end{itemize}

This large saxion condensate plays a crucial role in determining the dark matter abundance.  It eventually dominates the energy density of the universe and releases most of its entropy when it decays at temperature
\begin{equation}
T_{{\rm R}s} \simeq \sqrt{\Gamma_s M_{\rm Pl}}.
\label{eq:TRs}
\end{equation}
The decay rate of the saxion, $\Gamma_s$, is dependent on the saxion mass and on whether the Higgs doublets carry PQ charge.

In DFSZ models with $m_s >2 m_W$, the saxion mainly decays into a pair of Higgs, $W$ or $Z$ bosons, with the rate
\begin{align}
\Gamma_s(s \rightarrow hh,W^+W^-,ZZ) \; = \; \frac{q_\mu^2 \mu^4}{4\pi m_s V_{\rm PQ}^2 },
\label{eq:GammaSaxVisible}
\end{align}
where we have summed over the final states and assumed the decoupling and large ${\rm tan}\beta$ limits. The PQ charge $q_\mu$ of the Higgs mass parameter $\mu$ is normalized such that all charges of the PQ breaking fields are integers with absolute values as small as possible. We fix $q_\mu = 2$ in this paper, as in the minimal supersymmetric DFSZ model. For a lighter DFSZ saxion, $m_s<2 m_W$, the main decay channel is into a pair of standard model fermions via mixing with the Higgs, with the rate
\begin{align}
\Gamma_s(s \rightarrow f\bar{f}) = \frac{q_\mu^2}{4\pi} \frac{m_s\mu^4}{m_h^4 V_{\rm PQ}^2} \sum_{m_f<m_s/2} N_f m_f^2,
\label{eq:GammaSaxFermions}
\end{align}
where $N_f$ is the multiplicity of the fermion $f$ (3 and 1 for quarks and leptons, respectively). Here we have assumed the decoupling and large ${\rm tan}\beta$ limits as well as $m_s \ll m_h$.

In KSVZ models, the saxion mainly decays into a pair of gluons with a rate
\begin{align}
\Gamma_s(s\rightarrow gg) \; = \; \frac{\alpha_3^2}{32\pi^3} \frac{m_s^3}{f_a^2}.
\label{eq:GammaSaxGluons}
\end{align}
Here the axion decay constant $f_a$ is defined by the axion coupling with the gluon field as
\begin{align}
\mathcal{L}_{a G \tilde{G}} = \frac{g_s^2}{32 \pi^2} \frac{a}{f_a} G^{\mu\nu} \tilde{G}_{\mu\nu} \ ,
\end{align}
and is given by the PQ breaking scale $V_{\rm PQ}$ through the relation $f_a = \sqrt{2} V_{\rm PQ}/N_{\rm DW}$. 
In either case, $T_{{\rm R}s}$ is low enough that the gravitinos thermalized at early times are diluted.  For most values of $V_{\rm PQ}$, the decay of the saxion condensate also dilutes axinos and gravitinos from freeze-in (FI) and superpartners from freeze-out; and, for higher values of $V_{\rm PQ}$, even misalignment axions generated near the QCD phase transition are diluted.

In the next sub-section we discuss aspects of the saxion oscillation matter-dominated era.
In the following sub-section we provide a very simple illustration of a possible cosmology for the early evolution of the saxion condensate -- the ``Decoupled Saxion''. 

\subsection{The Matter Dominated Era of the Saxion Condensate}
In this paper the processes relevant for computing the dark matter abundance are dilution of previously thermalized gravitinos, freeze-in gravitino production and misalignment axions.  These processes all occur at low temperatures, $T \lsim \tilde{m}$.  Furthermore, constraints on the theory from overproduction of freeze-in axinos and superpartner freezeout similarly occur at $T \lsim \tilde{m}$.  Hence the results of this paper only depend on the low temperature aspects of the cosmology, not the high temperature aspects.  Thus the evolution of the saxion at $T > \tilde{m}$ could be arbitrarily complicated, for example from interactions during inflation or with the thermal bath.  Nevertheless, for the late evolution at $T \lsim \tilde{m}$ in the potential (\ref{eq:quadsaxpot}) we need to parameterize the size of the condensate.  

In particular well before it decays, the saxion condensate must dominate the energy density producing a matter dominated universe at $T \gg T_{{\rm R}s}$.  From $T_{{\rm R}s}$ up to some temperature $T_{\rm NA}$ this MD era is non-adiabatic (MD$_{\rm NA}$): the radiation density is dominated by the products of recently decayed saxions rather than from pre-existing red-shifted radiation, giving $T \propto 1/a^{3/8}$.  On the other hand at temperatures above $T_{\rm NA}$ there are so few saxion decays that the MD era is adiabatic (MD$_{\rm A}$), with $T \propto 1/a$.   At $T_{\rm NA}$ the saxion condensate has a size $s_{\rm NA} \simeq T_{\rm NA}^4/ T_{{\rm R}s}^2 m_s$ and we find it convenient to use $T_{\rm NA}$ to describe the strength of the saxion condensate, as it appears directly in the gravitino dilution factor.   For decoupled relic particles produced at temperatures above $T_{\rm NA}$, such as the previously thermalized gravitinos, saxion decays yield a dilution factor\footnote{If $T_{NA} > m_s$ this result gets corrected; however, we find $T_{NA} < m_s$ over a wide range of cases discussed below.} 
\begin{align}
D \simeq \left(  \frac{T_{\rm NA}}{T_{{\rm R}s}} \right)^5
\label{eq:D}
\end{align}
while for relic particles produced at some temperature $T$ between $T_{\rm NA}$ and $T_{{\rm R}s}$ the dilution factor is less
\begin{align}
D(T) \simeq \left(  \frac{T}{T_{{\rm R}s}} \right)^5.
\label{eq:DT}
\end{align}

The condition on $T_{\rm NA}$ follows from requiring that the dilution of previously thermalized gravitinos yields the observed temperature of matter-radiation equality
\begin{align}
T_{\rm eq}  \; \simeq \; \frac{m_{3/2} Y_{\rm th}}{ D}.
\label{eq:Teq}
\end{align}
We study gravitino dilution over a very wide range of parameters:  $V_{\rm PQ}$ is varied over its entire range from its lower astrophysical bound (see~\cite{Raffelt:2006cw} for a review) of $10^9 \, \GeV$ to $M_{\rm Pl}$, and the saxion mass is varied over the range of $10 \;{\rm MeV}< m_s < 10$ TeV.  Throughout this parameter space, the observed dark matter results from diluting previously thermalized gravitinos and/or gravitinos produced by freeze-in processes, and $T_{\rm NA}$ is constrained to be in the range 10 MeV $<T_{\rm NA} <$ 100 TeV.

In DFSZ theories with large $\mu$ and small $V_{\rm PQ}$, the formulae (\ref{eq:TRs}) and (\ref{eq:Teq}) give $T_{\rm NA},T_{{\rm R}s}>m_s$. For temperatures above $m_s$, however, the decay/scattering of the saxion is affected by thermal effects~\cite{Linde:1985gh,Yokoyama:2004pf,Drewes:2010pf}, which determine the temperatures $T_{{\rm R}s}$ and $T_{\rm NA}$.
For example, when the saxion has a Yukawa interaction $y s f\bar{f}$ with a fermion $f$, the decay (dissipation) rate of the saxion is given by $\Gamma\sim y^2 T$ for $T\gg m_s$. The resultant reheating temperature is given by $T_{{\rm R}s}\sim y^2 \mpl$, and the dilution factor is $D\sim (T_{\rm NA}/T_{{\rm R}s})^3$.
In the lower part of Figure~\ref{fig:gravMax}, thermal effects determine $T_{\rm Rs}$ and/or $T_{\rm NA}$. For simplicity we do not show the contours of required $T_{\rm NA}$ if $T_{\rm NA}>m_s$.  In Figure~\ref{fig:gravDFSZ} this thermal effect is irrelevant.

\subsection{Cosmology of the ``Decoupled Saxion''}
\label{subsec:DecpSax}

As a particular example of a saxion cosmology at $T > \tilde{m}$ we consider the ``Decoupled Saxion'', defined by the assumption that the saxion potential is given by Eq.~(\ref{eq:quadsaxpot}) for all temperatures back to 
\begin{align}
T_{\rm osc}  \simeq \sqrt{ m_s M_{\rm Pl}} \simeq 10^{10} \, \GeV  \left( \frac{m_s}{100 \GeV} \right)^{ \scalebox{1.01}{$\frac{1}{2}$} }.
\end{align}
This could happen if the saxion couples to the thermal bath via either very small dimensionless couplings or through suppressed higher dimension operators.  We stress that this is just a simple illustrative example, and is not necessary for the results of the next section.

Taking the reheat temperature after inflation, $T_R$, to be larger than $T_{\rm osc}$, the saxion field starts to oscillate at $T_{\rm osc}$ with some large amplitude $s_I$ that it acquired from some previous era, so that during the adiabatic era following $T_{\rm osc}$ the saxion energy density is
\begin{align}
\rho_s \simeq \frac{1}{2} \; m_s^2 s_I^2 \; \frac{T^3}{T_{\rm osc}^3}.
\end{align}
The universe becomes matter dominated by the saxion condensate at
\begin{align}
T_M \simeq \; \frac{m_s^2 s_I^2 }{T_{\rm osc}^3}
\end{align}
and the subsequent matter-dominated era becomes non-adiabatic at $T_{\rm NA}$, when the radiation bath becomes dominated by saxion decay products rather than by the red-shifted radiation from inflaton decay, with
\begin{align}
T_{{\rm NA}}^{5/2} \simeq m_s  \Gamma_s M_{\rm Pl} \; \frac{s_I}{T_{\rm osc}^{3/2}}.
\label{eq:TNA}
\end{align}
Hence, in this scenario it is best to describe the strength of the saxion condensate by $s_I$ and have $T_{\rm NA}$ as a derived quantity given by (\ref{eq:TNA}).  Furthermore, in this cosmology, $D$ of Eq.~(\ref{eq:D}) becomes $D= T_M/T_{{\rm R}s}$.
The gravitino dark matter abundance constraint of Eq.~(\ref{eq:Teq}) then leads to
\begin{align}
\left(\frac{s_I}{M_{\rm Pl}} \right)^2 \simeq \frac{m_{3/2} Y_{\rm th}}{T_{\rm eq}} \left( \frac{\Gamma_s}{m_s} \right)^{ \scalebox{1.01}{$\frac{1}{2}$} } 
\end{align}
which we will find yields very large values of $s_I$ in the range of $(10^{14}-10^{18}) \GeV$ and is correlated with other parameters according to
\begin{align}
s_I \simeq (10^{16}, 10^{17}) \, \GeV \;
\left( \left(\frac{m_s}{\rm{TeV}}\right)^{ \scalebox{1.01}{$\frac{1}{2}$} } , \left(\frac{\rm{TeV}} {m_s}\right)^{ \scalebox{1.01}{$\frac{1}{2}$} } \frac{\mu}{{\rm TeV}} \right)
\left( \frac{m_{3/2}}{\GeV} \; \frac{10^{12} \GeV}{V_{\rm PQ}} \; \right)^{ \scalebox{1.01}{$\frac{1}{2}$} }
\end{align}
for saxions decaying to pairs of (gluons, electroweak and Higgs bosons).

The required large field value may seem to be incompatible with the assumption of the quadratic potential.
Actually in models where the saxion mass is generated via quantum corrections, the saxion potential in general becomes logarithmic at a large field value, and hence the assumption breaks down. However, in models where the saxion mass is given by a tree-level superpotential~\cite{Carpenter:2009zs,Carpenter:2009sw,Harigaya:2017dgd}, the saxion potential may be quadratic even for a large field value.

\subsection{Misalignment Axion Contribution to Dark Matter}

The axion is produced by the usual misalignment mechanism but now we need to consider the effect of saxion dilution~\cite{Lazarides:1987zf,Kawasaki:1995vt,Hashimoto:1998ua}. The axion field value is initially displaced from the minimum today by an amount $f_a \theta_i$, where $\theta_i$ is the misalignment angle.  Coherent oscillations of the axion field commence at temperature $T_{\rm osc}^{(a)}$, when the Hubble rate is equal to the axion mass.
We assume that the axion oscillation starts during the MD$_{\rm NA}$ era, which is the case for the parameter spaces constrained by the axion abundance.
In this case, the calculation is independent of $T_{\rm NA}$. We evaluate the axion energy density per entropy density at the end of entropy production~\cite{Co:2016vsi}
\begin{align}
\label{eq:AxionAbundance}
\left. \frac{\rho_{a}}{s}  \right|_{T_{{\rm R}s}} = \frac{9}{8}  \frac{f_a^2 \theta_{i}^2}{ M_{\rm Pl}^2}  \frac{T_{{\rm R}s}}{\xi(T_{{\rm R}s})} ,
\end{align}
where $\xi(T) \equiv m_a(T_{\rm osc}^{(a)})/m_a(T) (\le 1$ for $T<T_{\rm osc}^{(a)}$)  takes into account the temperature dependence of the axion mass.
We assume a simple power law $m_a^2(T) = m_a^2(0) (\Lambda/T)^\gamma$ above the QCD scale $\Lambda$.
The mass takes a constant value, $m_a(0) = 6 \, \EV \, (10^6 \GEV / f_a)$, at $T<\Lambda$. In other words, if $T_{\rm osc}^{(a)} \le \Lambda$, $\xi = 1$; otherwise, $\xi = (\Lambda/T_{\rm osc}^{(a)})^{\gamma/2}$ is used to compute the axion energy density today. The analytic formula of $\xi$ was derived in Ref.~\cite{Co:2016vsi}.

\begin{figure}[tb]
 \begin{center}
  \includegraphics[width=0.6\linewidth]{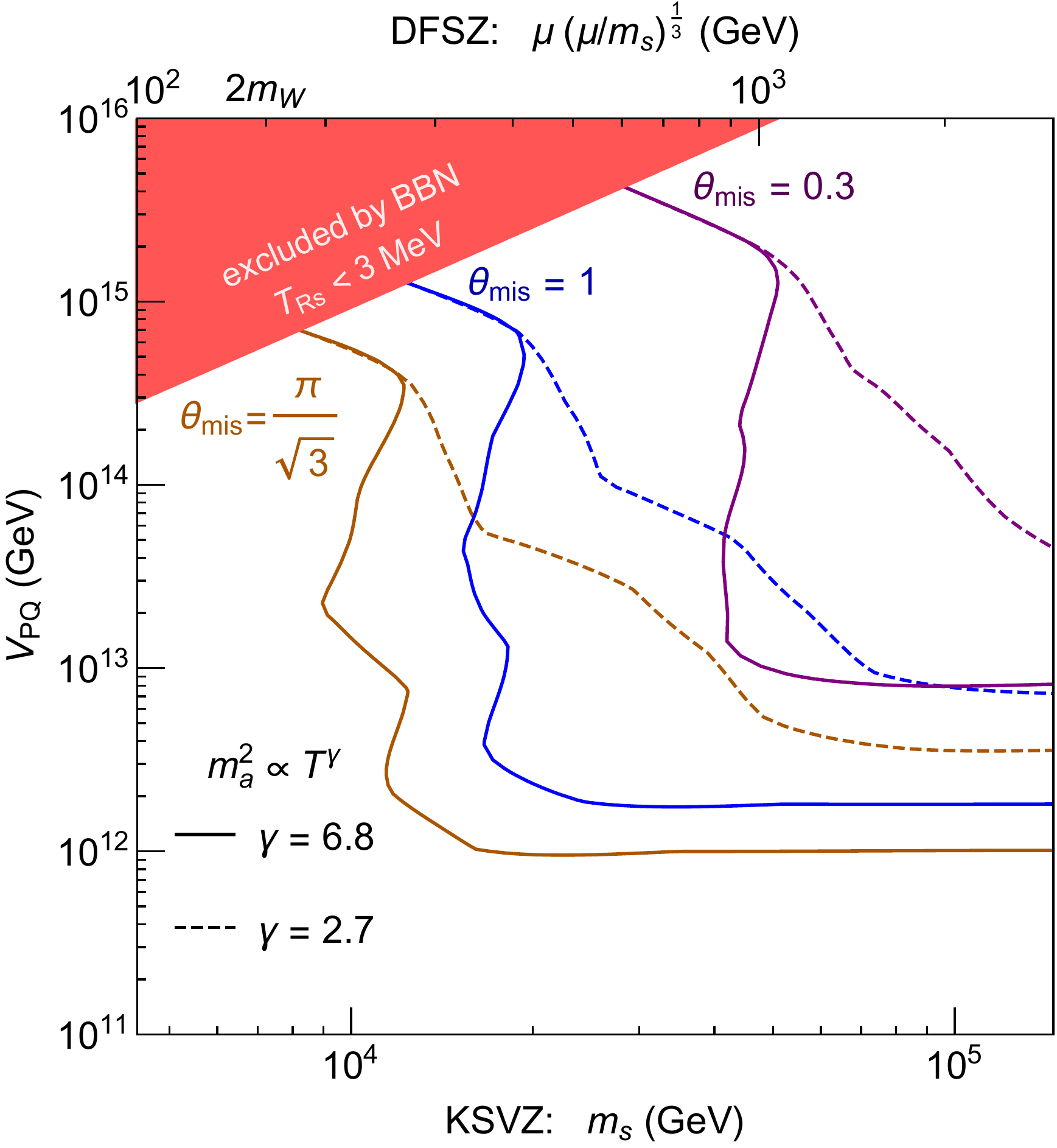}
 \end{center}
\caption{\small\sl 
Contours of the axion abundance, $\Omega_a h^2 = 0.11$, from vacuum misalignment and dilution from decay of the saxion condensate in DFSZ with $m_s>2m_h$ (upper axis) and KSVZ (lower axis) theories. Here we set $N_{\rm DW}=1$.
}
\label{fig:axion}
\end{figure}

We predict the axion abundance in terms of $T_{{\rm R}s}$ and $f_a = \sqrt{2} V_{\rm PQ}/N_{\rm DW}$. We use $N_{\rm DW} = 1$, while $T_{{\rm R}s}$ in Eq.~(\ref{eq:TRs}) can be calculated for both DFSZ and KSVZ theories. Figure~\ref{fig:axion} shows the numerical result of the contours of $\Omega_ah^2 = 0.11$ for various misalignment angles $\theta_i$ and the axion mass index $\gamma$ obtained from lattice calculations.
The region above and to the right of the contour is excluded by axion overproduction. The top (bottom) axis refers to the set of parameters necessary to compute $T_{{\rm R}s}$ in DFSZ (KSVZ) theories. For DFSZ, we assume  $m_s$ in a range where the saxion decays dominantly to $W^+W^-$, $ZZ$ and $hh$. The solid (dashed) lines are for the axion mass index $\gamma=6.8$ (2.7) computed in Ref.~\cite{Borsanyi:2015cka}(\cite{Bonati:2015vqz}). In the regions where the two index lines merge, the axion starts oscillating after its mass is already a constant, i.e. $T_{\rm osc}^{(a)}<\Lambda$. Recent lattice calculations   \cite{Petreczky:2016vrs, Frison:2016vuc, Borsanyi:2016ksw, Taniguchi:2016tjc} show that the axion mass index is well described by the dilute instanton gas approximation, $\gamma\simeq8$, in high temperature regimes. The red region is excluded by BBN because of the late decays of the saxion.

Interestingly, including the dependence of $\xi$ on $V_{\rm PQ}$, one finds that $\Omega_a h^2$ decreases (increases) for $\gamma>4$ ($\gamma<4$) when $V_{\rm PQ}$ increases. This explains the different overall slopes of the two index lines, while the detailed features arise due to the rapid change of $g_*(T)$ during the QCD phase transition. In the case of $\gamma = 6.8$, the dependence on $V_{\rm PQ}$ and effects of $g_*(T)$ compete with each other, resulting in a nearly vertical contour. We compute $g_*(T)$ from the Boltzmann distribution with the full SM spectrum and the MSSM spectrum degenerate at $1 \, \TEV$, and we linearly interpolate $g_*(T)$ across the QCD phase transition, i.e. $100 \, \MEV < T < 300 \, \MEV$.

The axion abundance gives an upper bound on the higgsino mass in the DFSZ model.
The bound is stringent for $m_s>2m_W$.
Note first that the saxion mass cannot be larger than $2\mu$, since otherwise saxions decay into pairs of higgsinos and result in too large gravitino abundance.
Then from the upper x-axis of Figure~\ref{fig:axion}, we can derive an upper bound on $2^{-1/3} \mu\simeq 0.8 \mu$. 
Assuming the the fine-tuning in the misalignment angle is no more than 10\%, the higgsino mass should be smaller than about $1$ TeV for $V_{\rm PQ}\gsim 10^{13}$ GeV.
If the misalignment angle takes a randomized value, the axion abundance should be evaluated with the averaged angle, $\theta_{\rm mis} \simeq \pi/ \sqrt{3}$. Then the higgsino mass should be smaller than about $200$ GeV for $V_{\rm PQ}\gsim 10^{12}$ GeV.
This includes the cases where the PQ symmetry is unbroken during inflation, restored after inflation, or the axion field obtains large fluctuation due to the parametric resonance effect from saxion oscillations. The last case is discussed in Sec.~\ref{subsec:parametric resonance}.
For $m_s<2m_W$, the upper bound is relaxed.
Since a fermion with a mass close to $m_s$ dominantly contributes to the decay rate in Eq.~(\ref{eq:GammaSaxFermions}),
the bound is relaxed roughly by a factor of $(100~{\rm GeV}/m_s)^{3/4}$ in comparison to the case with $m_s>2m_W$.

\section{Thermalized Gravitino Dark Matter}

In this section we show that dilution by the late decay of a saxion condensate allows the observed dark matter abundance to arise from the thermalized gravitinos of an early epoch over a very wide range of $m_{3/2}$ and $V_{\rm PQ}$.  However, gravitinos can be overproduced by reactions occurring at the TeV scale or below: gravitino freeze-in, axino freeze-in and decay to gravitinos, and saxion decays to $\tilde{a} + \tilde{G}$.  We illustrate how these constrain the region where dark matter arises from the primordially thermalized gravitinos.  Similarly, we indicate where axions are overproduced, by either early misalignment or parametric resonance during saxion oscillations.  

There are several relevant parameters.  We show results for essentially complete ranges of $(m_{3/2}, V_{\rm PQ})$, but choose a few illustrative values for the key parameters $(m_s, m_{\tilde{a}}, \mu)$ and for other supersymmetry breaking parameters.  Our aim is not to provide an exhaustive study of the $(m_s, m_{\tilde{a}}, \mu)$ space, but to illustrate the wide range that allows thermalized gravitino dark matter and its corresponding rich signals.  
We examine constraints on the parameter space from other processes creating gravitinos and axions in sub-section B (C) for dominant saxion decays to gluons (Higgs/electroweak bosons). 

We comment on the lower bound on the saxion mass.
In KSVZ theories, in order for the saxion to decay before the BBN,
\begin{align}
m_s > 0.6~{\rm GeV} \times \left(\frac{f_a}{8\times 10^8~{\rm GeV}}\right)^{2/3}~~~~\text{KSVZ}
\end{align}
is required.
In DFSZ theories the saxion mass may be smaller due to its effective couplings with standard model fermions through its mixing with the Higgs.
However, with such a mixing, the saxion takes away energy from supernovae and changes the duration of the neutrino emission~\cite{Ellis:1987pk,Raffelt:1987yt,Turner:1987by,Mayle:1987as}.
To prevent this process requires
\begin{align}
m_s > {\cal O}(10)~{\rm MeV}~~~~\text{DFSZ}.
\end{align}

\subsection{The Maximal Parameter Space}
\label{subsec:Max}

\begin{figure}
\begin{center}
\includegraphics[width=0.495\linewidth]{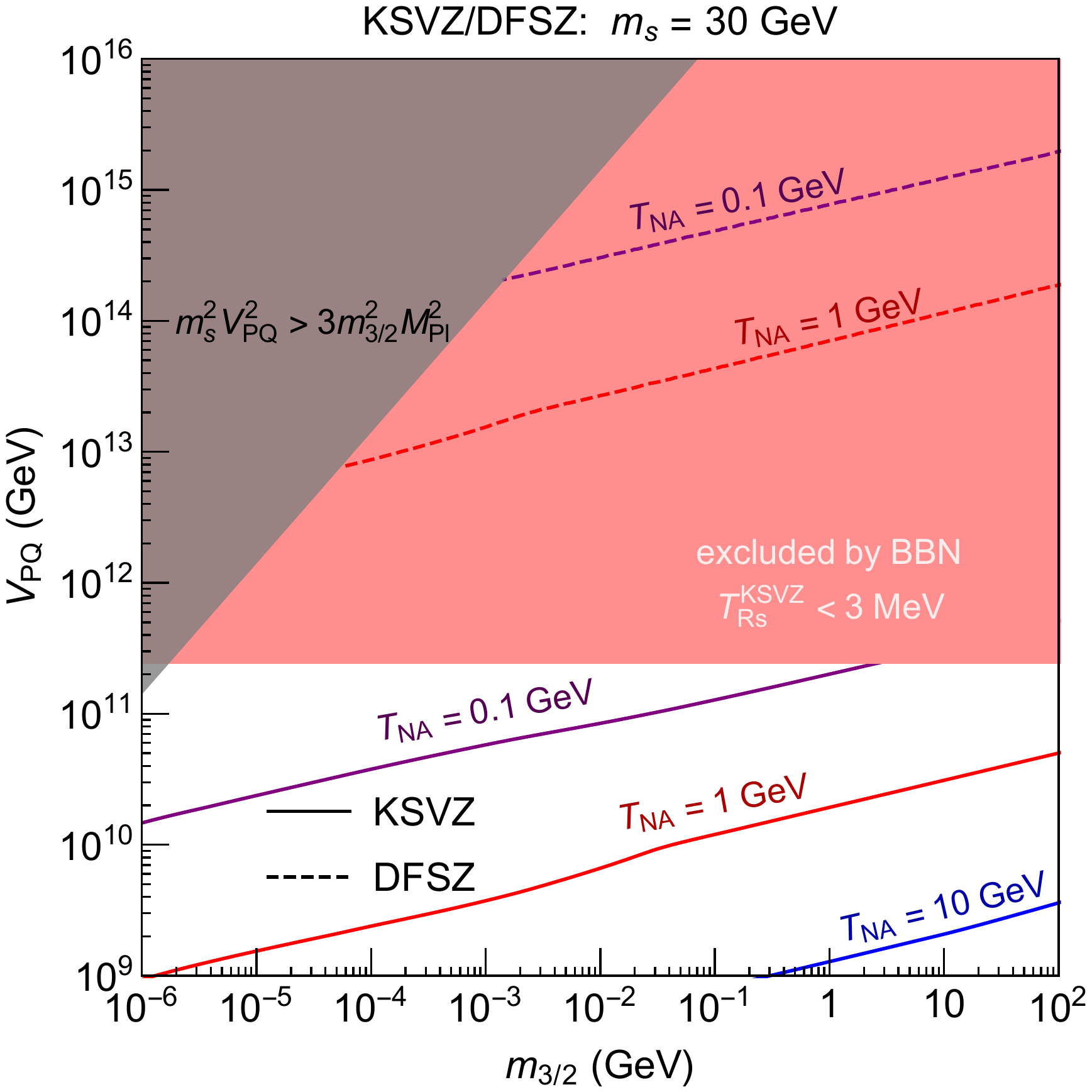} \includegraphics[width=0.495\linewidth]{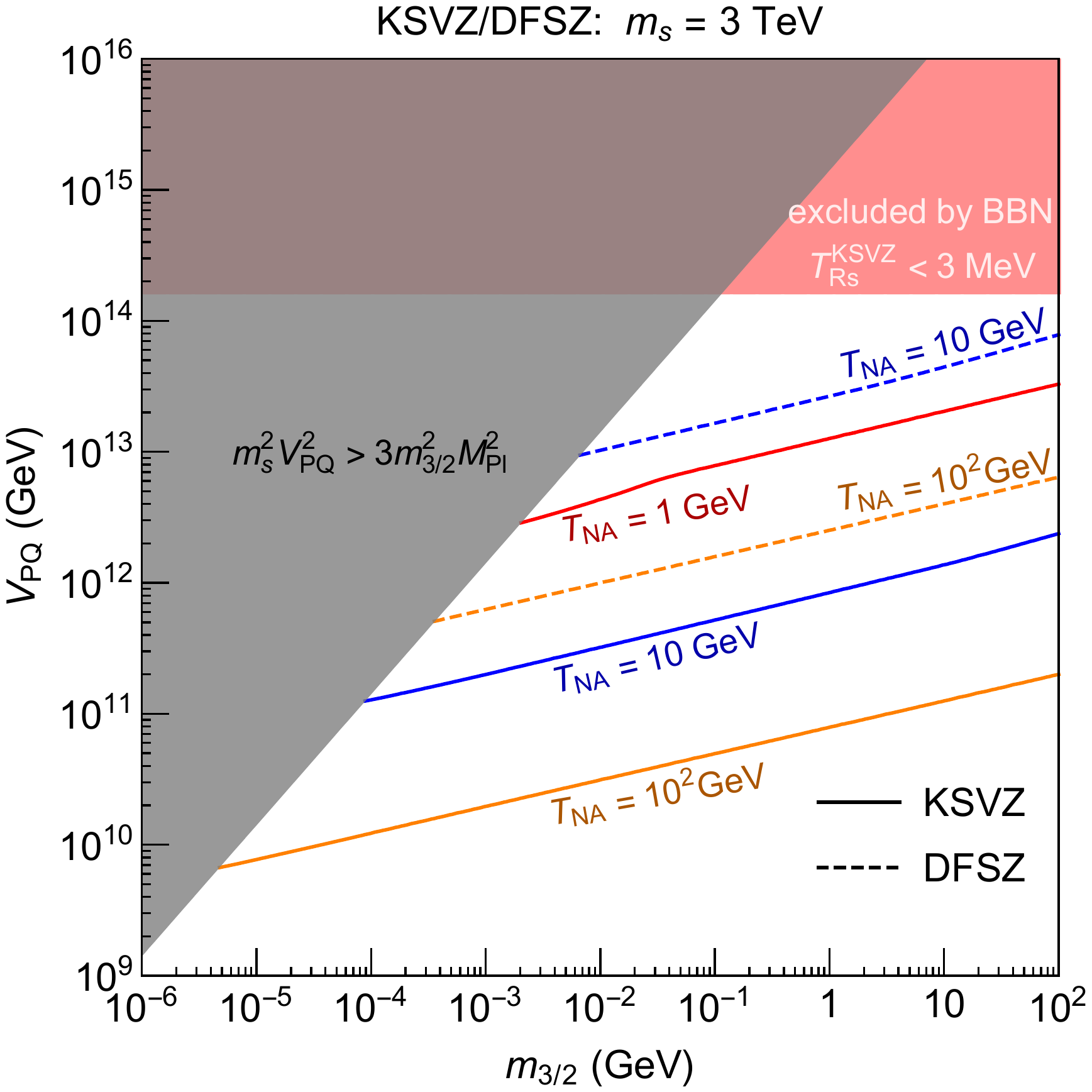}
\end{center}
\caption{The maximal parameter space for thermalized gravitino dark matter.  In the left panel, $m_s = 30$ GeV and saxion decay is dominated by $s \rightarrow gg$ ($s \rightarrow b\bar{b}$) for KSVZ (DFSZ with $\mu = 4$ TeV).  In the right panel $m_s = 3$ TeV and in KSVZ theories $s \rightarrow gg$ dominates while in DSFZ theories with $\mu=2$ TeV $s \rightarrow WW, ZZ, hh$ dominates. For both panels, the red region excluded by BBN applies only to KSVZ theories. We set $N_{\rm DW}=1$ for KSVZ.}
\label{fig:gravMax}
\end{figure}

The unshaded regions of Figure \ref{fig:gravMax} show the maximum ranges of $(m_{3/2}, V_{\rm PQ})$ that allow for thermalized gravitino dark matter from saxion condensate dilution. The left/right panel is for $m_s = 30$ GeV/ 3 TeV and each panel applies to both KSVZ and DFSZ theories.  Values of $m_{3/2}$ below $\approx$ keV are excluded by warm dark matter constraints, while values above 100 GeV  are possible as long as the gravitino remains the LSP.

The gray shaded region of Figures \ref{fig:gravMax}, \ref{fig:gravKSVZ} and \ref{fig:gravDFSZ} is excluded because the contribution to the vacuum energy from the saxion potential, of order $m_s^2 V_{\rm PQ}^2$, exceeds that allowed by total supersymmetry breaking $F_{\rm tot}^2 = 3 \, m_{3/2}^2 M_{\rm Pl}^2$.  This bound on $V_{\rm PQ}$ scales as $m_{3/2}/m_s$.  The bound is saturated in models where supersymmetry and PQ symmetry are simultaneously broken, and the saxion obtains its mass at tree level~\cite{Carpenter:2009zs,Carpenter:2009sw,Harigaya:2017dgd}.
In models where the saxion mass is given by quantum corrections, the bound is stronger by coupling constants and associated loop factors.

The red shaded region of Figures \ref{fig:gravMax}, \ref{fig:gravKSVZ} and \ref{fig:gravDFSZ} is excluded because the reheat temperature from saxion decays, $T_{{\rm R}s}$, is below 3 MeV destroying the success of BBN \cite{Kawasaki:2004yh}.  In KSVZ theories, where the dominant saxion decay is $s \rightarrow gg$, this bound on $V_{\rm PQ}$ scales as $m_s^{3/2}$.  In both panels of Figure \ref{fig:gravMax}, the red shading applies only to KSVZ theories. For the right panel in DFSZ theories, the dominant saxion decay is $s \rightarrow WW, ZZ, hh$ giving a bound on $V_{\rm PQ}$ that scales as $\mu^2/m_s^{1/2}$.  We have taken $\mu = 2$ TeV so that this bound on $V_{\rm PQ}$ is larger than $10^{16}$ GeV and does not appear in the figure.  In the left panel for DFSZ theories, the dominant saxion decay is $s \rightarrow \bar{b} b$ giving a bound on $V_{\rm PQ}$ that scales as $\mu^2 m_s^{1/2}$ and again is larger than $10^{16}$ GeV and does not appear.

The saxion decays into a pair of gravitinos through its mixing with the sgoldstino field or the mixing of the axino with the gravitino. The decay rate of the saxion into a pair of gravitinos is
\begin{align}
\Gamma (s \rightarrow \tilde{G} \tilde{G}) = \frac{\kappa{'^2}}{288\pi} \left( \frac{V_{\rm PQ}}{\mpl} \right)^2 \frac{m_s^5}{ m_{3/2}^2 \mpl^2}.
\end{align}
Here $\kappa'$ is an $O(1)$ parameter which depends on the couplings between the PQ breaking field and the supersymmetry breaking field, and may be suppressed if there is an (approximate) $Z_2$ symmetry in the couplings.
Even if $\kappa'={\cal O}(1)$, we found that this decay mode does not give additional constraints beyond the gray and red shaded regions.

These bounds, leading to the gray and red excluded regions, are inherent to the saxion condensate dilution mechanism and cannot be evaded.  Other processes producing gravitinos or axions are frequently important, and may lead to further constraints in the $(m_{3/2}, V_{\rm PQ})$ plane, but they depend on other parameters, such as the axino mass, or on details of cosmological evolution at temperatures far above the TeV scale.  Hence we omit them from Figure \ref{fig:gravMax}, which shows the maximal allowed region, and consider them at length in the next two sub-sections and in Figure~\ref{fig:gravKSVZ} and Figure~\ref{fig:gravDFSZ}.  Here we provide a brief qualitative illustration of how these other constraints can be avoided.

Contributions to gravitino dark matter from $s \rightarrow \tilde{G} \tilde{a}, \tilde{a}\tilde{a}$ are avoided by taking $m_{\tilde{a}} > m_s$.  A sufficiently large $m_{\tilde{a}}$ also removes constraints from axino freeze-in.  Effects on BBN arising from the lightest observable supersymmetric particle (LOSP) freezeout and decay can be made sufficiently mild by having a sneutrino LOSP.  With a sneutrino mass of 300 GeV the freezeout abundance is quite small; and neutrinos from decay to $\nu \tilde{G}$ have only mild effects on BBN \cite{Kawasaki:2008qe}.
A stau LOSP also makes the BBN constraint mild; a gravitino mass below 10 GeV is allowed.
The gravitino freeze-in abundance is controlled by the LOSP mass, and 300 GeV is already large enough to provide a sub-dominant contribution to dark matter.  A crucial feature of our scheme is that freeze-in of both axinos and gravitinos are highly suppressed as they occur during a matter dominated era and are subsequently diluted by saxion decays.  We also note that in KSVZ theories the decay $s \rightarrow aa$ must be mildly suppressed for $s \rightarrow gg$ to dominate.  Finally, there is the possibility that during the oscillation of the saxion field inhomogeneities in the axion field are exponentially enhanced by parametric resonance.   However, the importance of this effect depends on the very early cosmological evolution of the saxion field, and is model dependent.  While all these constraints can be avoided, they are frequently important and we discuss them quantitatively below in Secs.~\ref{subsec:gg} and \ref{subsec:WW}.

We find that previously thermalized gravitinos decouple from the thermal bath at a temperature higher than $T_{\rm NA}$. This means a large amount of entropy is injected only after these gravitinos stop interacting with the bath. As a result, the gravitino abundance is diluted by the factor $D$ of Eq.~(\ref{eq:D}).  Requiring dilution to yield the observed dark matter abundance via Eq.~(\ref{eq:Teq}), gives an analytic estimate for $T_{\rm NA}$.  In KSVZ theories saxions decay dominantly to gluons, with a rate given in Eq.~(\ref{eq:GammaSaxGluons}), giving 
\begin{align}
\label{eq:TNAKSVZ}
T_{\rm NA} (s \rightarrow gg) \simeq 100 \text{ MeV} \,  \left( \frac{m_s}{100 \text{ GeV}} \right)^{ \scalebox{1.01}{$\frac{3}{2}$} }  \left( \frac{10^{12} \ \text{GeV}}{V_{\rm PQ}} \right)   \left( \frac{m_{3/2}}{\text{MeV}} \right)^{ \scalebox{1.01}{$\frac{1}{5}$} } \; .
\end{align}
In DFSZ theories with $m_s > 2 \, m_W$, the decay rate of Eq.~(\ref{eq:GammaSaxVisible}) gives
\begin{align}
T_{\rm NA} (s \rightarrow WW, ZZ, hh) \simeq 100 \text{ GeV} \,  \left( \frac{\mu}{\text{TeV}} \right)^{ \scalebox{1.01}{$\frac{3}{2}$} } \left( \frac{\mu}{m_s} \right)^{ \scalebox{1.01}{$\frac{1}{2}$} }   \left( \frac{10^{12} \ \text{GeV}}{V_{\rm PQ}} \right)   \left( \frac{m_{3/2}}{\text{MeV}} \right)^{ \scalebox{1.01}{$\frac{1}{5}$} } \; ,
\end{align}
while for $m_s < 2 \, m_W$ the decay rate in Eq.~(\ref{eq:GammaSaxFermions}) gives
\begin{align}
T_{\rm NA} (s \rightarrow f \bar{f}) \simeq 20 \text{ GeV} \,  \left( \frac{\mu}{\text{TeV}} \right)^{ 2 } \left( \frac{m_s}{100 \text{ GeV}} \right)^{ \scalebox{1.01}{$\frac{1}{2}$} }   \left( \frac{10^{12} \ \text{GeV}}{V_{\rm PQ}} \right)   \left( \frac{m_{3/2}}{\text{MeV}} \right)^{ \scalebox{1.01}{$\frac{1}{5}$} } \left( \frac{N_f}{3} \right)^{ \scalebox{1.01}{$\frac{1}{2}$} } \frac{m_f}{m_b} \; .
\end{align}
In all cases we take $g_*(T_{{\rm R}s}) = 10.75$.

Numerical results for $T_{\rm NA}$ are shown in Figure~\ref{fig:gravMax} in the $(m_{3/2}, V_{\rm PQ})$ plane as solid/dashed contours for saxion decays to $gg/(WW, ZZ, hh)$.  Throughout the entire allowed region $T_{\rm NA} < \cal{O}$ (TeV) so that dilution of previously thermalized gravitinos by decay of the saxion condensate involves cosmology of the TeV era or later.  Hence we are able to discuss this scenario in a very general framework, without the need to specify a particular UV theory, by making the two key assumptions listed at the beginning of section II.

\subsection{Further Constraints in KSVZ Theories with $s \rightarrow gg$ }
\label{subsec:gg}

\begin{figure}
\begin{center}
\includegraphics[width=0.495\linewidth]{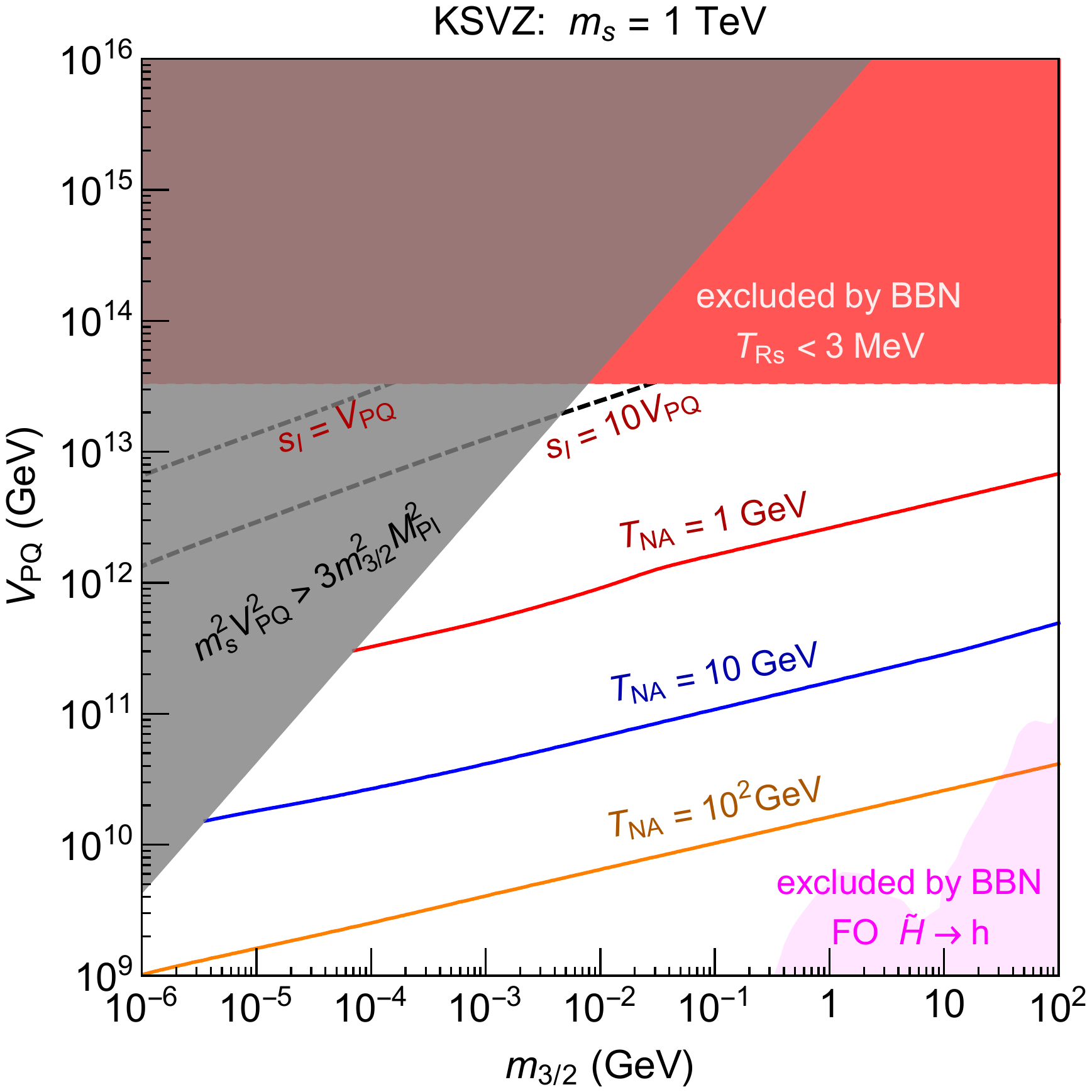} \includegraphics[width=0.495\linewidth]{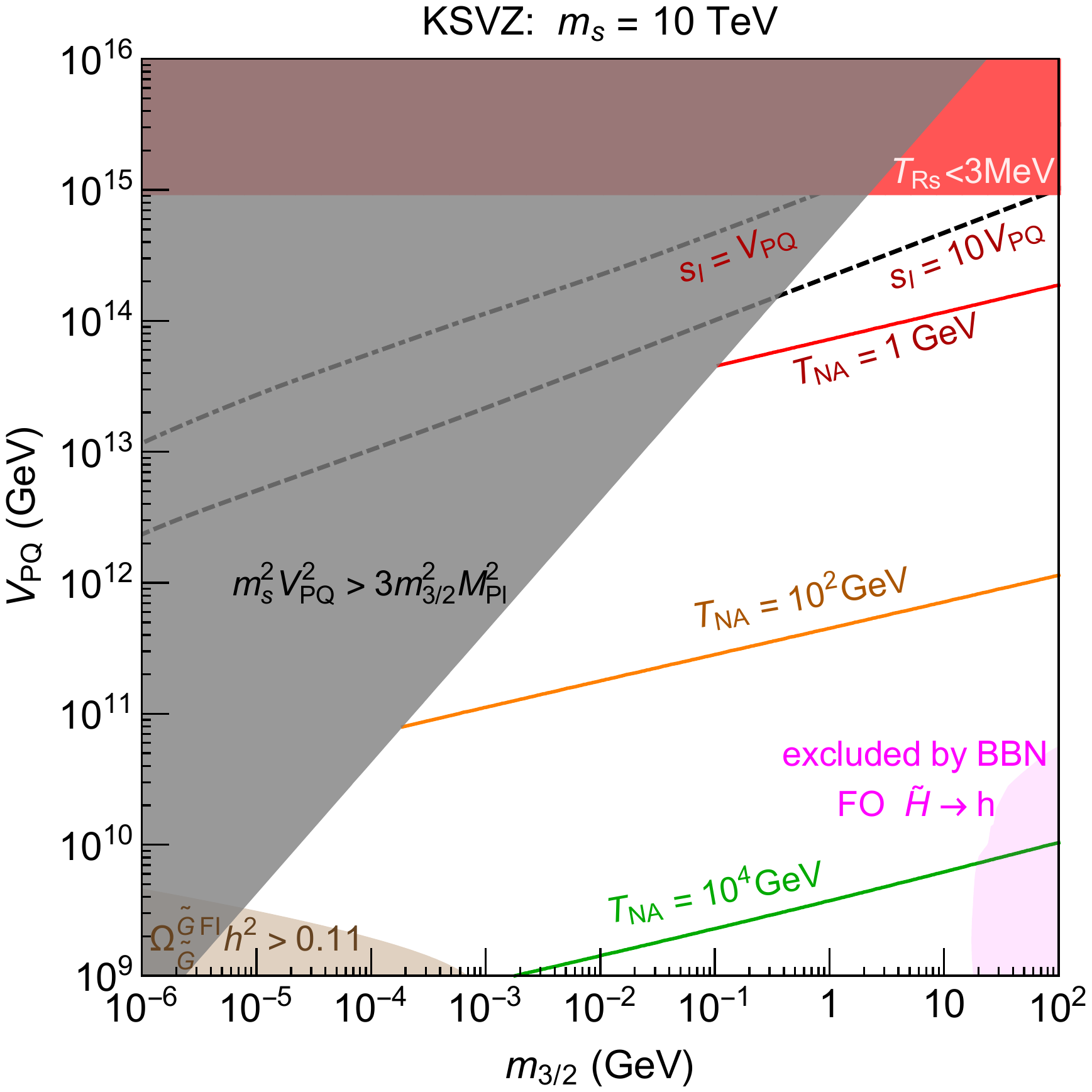}
\end{center}
\caption{Constraints on gravitino dark matter in KSVZ theories, with saxions decaying dominantly to gluons. In deriving the BBN constraint on freeze-out higgsinos, we take $M_{1,2} \gg \mu =$ 1 TeV (6 TeV) for the left (right) panel. We set $N_{\rm DW}=1$.}
\label{fig:gravKSVZ}
\end{figure}

We explore further constraints on KSVZ theories from freeze-in of gravitinos and freeze-out of the LOSP.  First we perform a similar calculation of $T_{\rm NA}$ contours as in Figure~\ref{fig:gravMax}, with numerical results given in Figure~\ref{fig:gravKSVZ} for $m_s = (1,10)$ TeV in the (left, right) panel. The allowed parameter space is the white (unshaded) region. The red and gray regions excluded by BBN and the consistency of supersymmetry breaking are discussed in Sec.~\ref{subsec:Max}. The dot-dashed (dashed) lines of Figure~\ref{fig:gravKSVZ} and Figure~\ref{fig:gravDFSZ} give contours of $s_I = V_{\rm PQ} \ (10\, V_{\rm PQ})$, using Eq.~(\ref{eq:TNA}).   $s_I$ is the initial saxion field value in the ``Decoupled Saxion" cosmology of Sec.~\ref{subsec:DecpSax} and the importance of these contours for axion parametric resonance is discussed in Sec.~\ref{subsec:parametric resonance}.

In addition to gravitinos thermalized during an early epoch, the decays of supersymmetric particles to gravitinos also contributes to the final abundance via the FI mechanism.  The freeze-in process is IR dominated and terminates when the abundance of supersymmetric partners becomes exponentially suppressed as the temperature falls below their masses.   Since gravitinos interact with all multiplets via the goldstino interaction, this FI contribution is determined mainly by the LOSP and hence the LOSP mass. For illustration purposes, we assume a higgsino LOSP.

The FI abundance is proportional to the decay rate, which is enhanced for low gravitino masses and high parent particle masses. As shown in Figure~\ref{fig:gravKSVZ}, the FI contribution is absent for $\mu = 1~\TEV$ because the higgsino LOSP decay is inefficient. However, with $\mu = 6~\TEV$ in the right panel, freeze-in gravitinos become the dominant source in the brown shaded region. Furthermore, since the freeze-in of gravitinos occurs below $T_{\rm NA}$ in this region, the dilution factor is given by Eq.~(\ref{eq:DT}) and depends only on $\mu$ and $T_{{\rm R}s}$ but not $T_{\rm NA}$. As a result, the brown region is excluded because the saxion decays too early to provide sufficient dilution for FI gravitinos, regardless of $T_{\rm NA}$.

The regions shaded in magenta in Figure~\ref{fig:gravKSVZ} are excluded by BBN due to late decays of higgsinos $\tilde{H}$ produced via freeze-out. The subsequent decays of higgsinos place constraints on high gravitino masses, where the decay is late, and on low $V_{\rm PQ}$, where the saxions decay early.   The higgsino mass, $\mu$, is increased from the left to the right panel, increases both the higgsino decay rate and the higgsino FO abundance.  The BBN constraint shifts to the right and upward. 

\subsection{Further Constraints in DFSZ Theories with $s \rightarrow hh/ZZ/WW$}
\label{subsec:WW}

\begin{figure}[t]
 \begin{center}
  \includegraphics[width=0.485\linewidth]{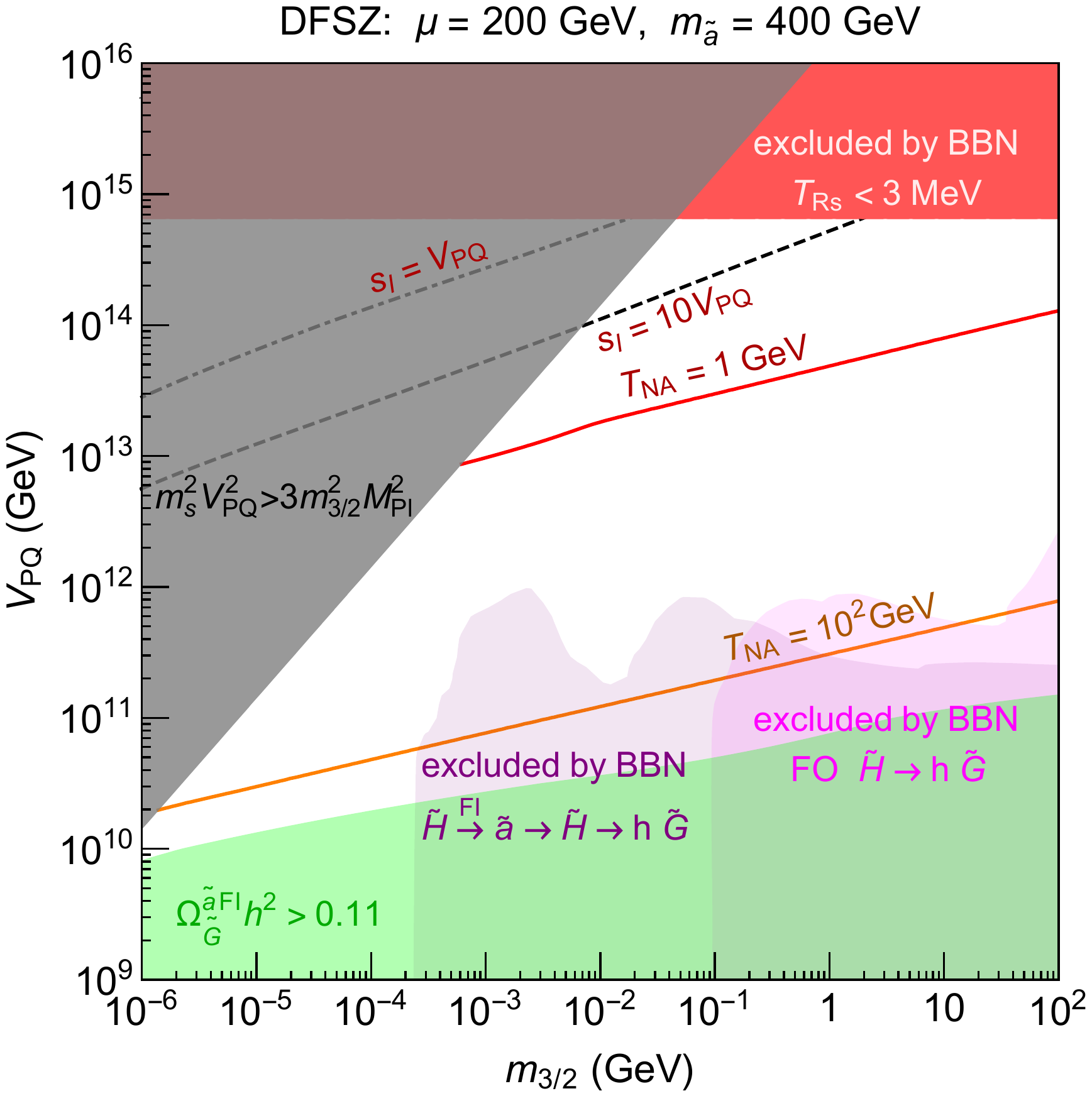}  \includegraphics[width=0.485\linewidth]{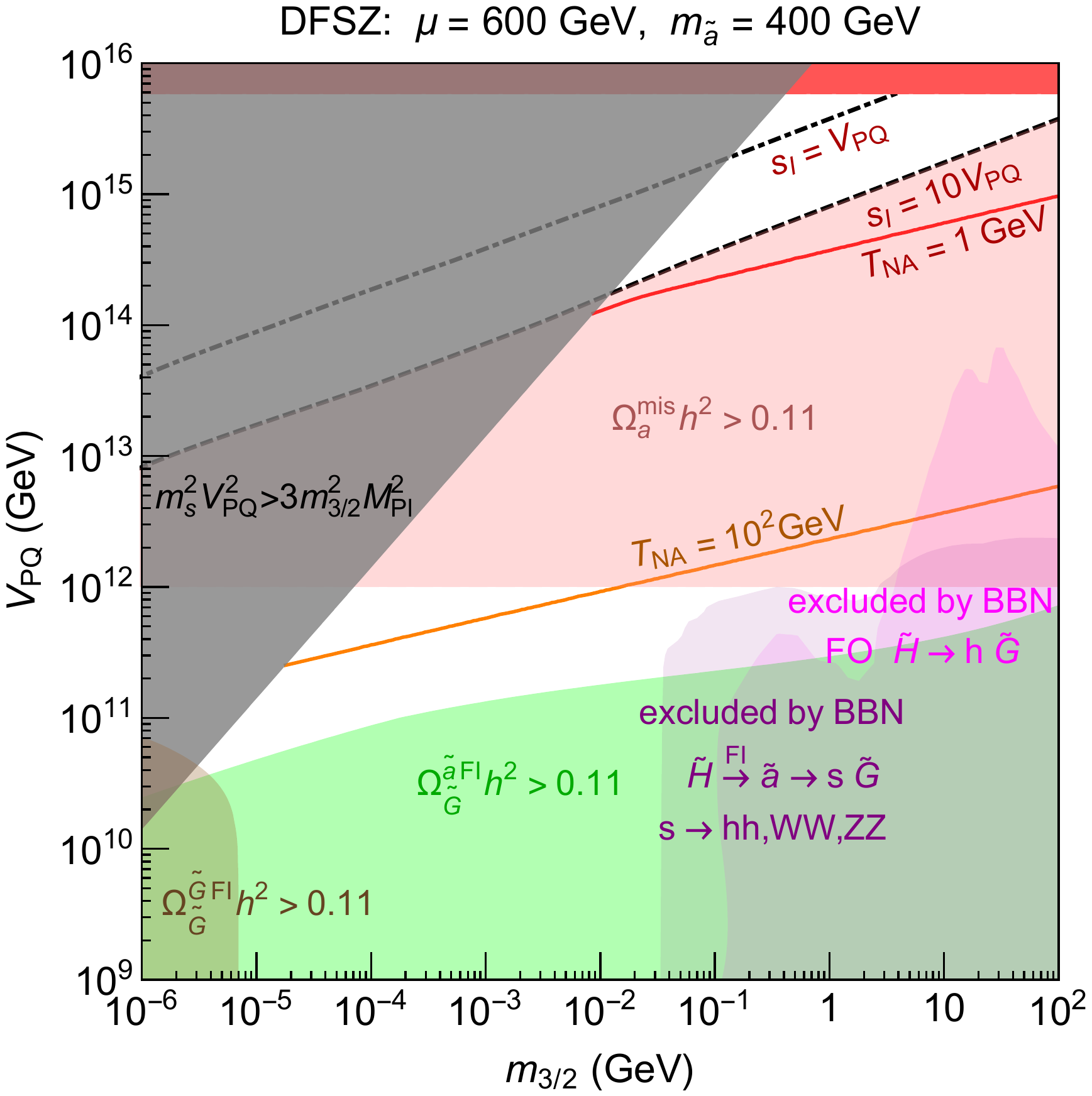} \\ \vspace{0.25cm}
  \includegraphics[width=0.485\linewidth]{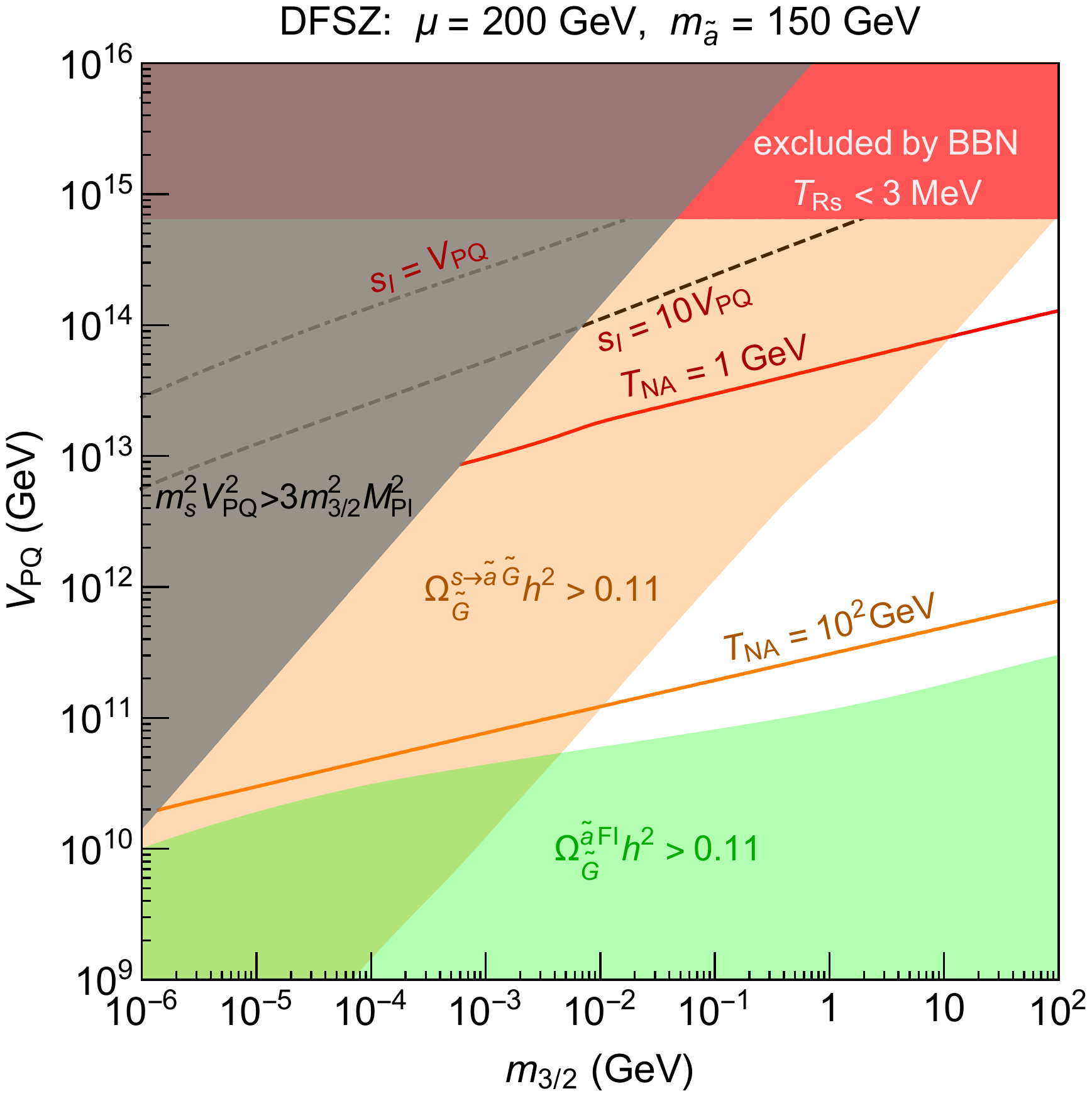}  \includegraphics[width=0.485\linewidth]{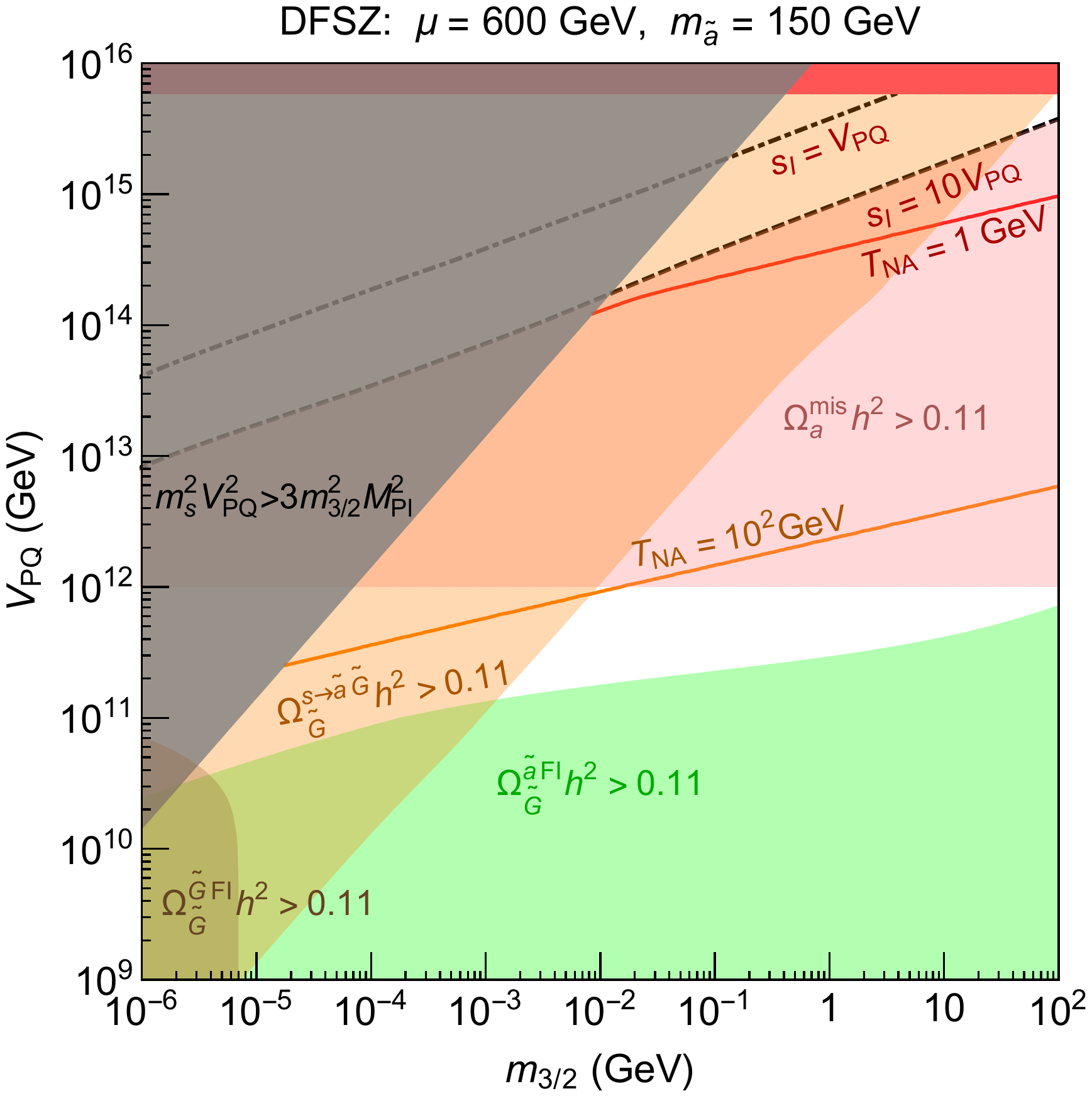}
 \end{center}
\caption{
Constraints on gravitino dark matter in DFSZ theories with saxions of mass $m_s = 300$ GeV decaying dominantly to Higgs and electroweak bosons, for $q_\mu = 2$.  For the left (right) two panels $\mu =200$ GeV (600 GeV) and we take $M_{1,2} \gg \mu$. In the upper (lower) two panels we take $m_{\tilde{a}} = 400 \ \GeV > m_s \ (m_{\tilde{a}} = 150 \ \GeV < m_s$). The axion abundance is computed with $N_{\rm DW}=1$.
}
\label{fig:gravDFSZ}
\end{figure}

The analysis for KSVZ and DFSZ theories has both similarities and differences, as seen by comparing Figures \ref{fig:gravKSVZ} and \ref{fig:gravDFSZ}. In Figure \ref{fig:gravDFSZ} we fix $m_s = 300$ GeV, while $\mu$ and $m_{\tilde{a}}$ take on different values in the four panels, such that in the upper (lower) panels, $m_{\tilde{a}} > m_s$ ($m_{\tilde{a}} < m_s$).  Numerical solutions for contours of $T_{\rm NA}$ are given in the $(m_{3/2}, V_{\rm PQ})$ plane.   The white (unshaded) regions give the allowed parameter space, which is significantly limited by several constraints as discussed below.

While the gray shaded regions requiring $m_{3/2} V_{\rm PQ} < \sqrt{3} F_{\rm tot}$ are the same as in KSVZ theories, the red shaded region from BBN limits on saxion decay are much milder,  especially for $m_s > 2m_W$.   
The saxion decay rate to gluons, Eq.~(\ref{eq:GammaSaxGluons}), is loop-suppressed relative to the decay rate to electroweak bosons, Eq.~(\ref{eq:GammaSaxVisible}). In DFSZ theories, this faster saxion decay rate increases $T_{{\rm R}s}$ so that the region with $T_{{\rm R}s} < 3$ MeV excludes only the largest values of $V_{\rm PQ}$.

As in the KSVZ case, DFSZ theories can have small regions at low $V_{\rm PQ}$ excluded by over-closure from gravitino freeze-in (brown region at low $m_{3/2}$) and from BBN limits from the decays of freeze-out higgsinos, $\tilde{H} \rightarrow h \tilde{G}$, (magenta region at large $m_{3/2}$).   The former disappears in the left panels because the higgsino decay rate to gravitinos depends strongly on $\mu$. The latter disappears in the lower panels because the higgsinos from FO decay before BBN to axinos which then decay harmlessly to $a \tilde{G}$.

A key feature of DFSZ theories is additional production processes involving axinos, if $m_{\tilde{a}}$ is not too large, giving the large green, purple and orange regions of Figure~\ref{fig:gravDFSZ}.

Axino FI and decay to gravitinos yields too much dark matter in the green regions at low $V_{\rm PQ}$. The freeze-in of axinos (and gravitinos) from decays of the higgsino LOSP occurs during the MD$_{\rm NA}$ era or after saxion reheating for low enough $V_{\rm PQ}$ and $m_{3/2}$. This implies that the relevant dilution factor is (\ref{eq:DT}) and is insensitive to $T_{\rm NA}$ and solely determined by the freeze-in temperature and $T_{{\rm R}s}$. 

Axino FI and decay is excluded by BBN in the purple regions of Figure~\ref{fig:gravDFSZ}.  In the top left panel the relevant decay chain is $\tilde{a} \rightarrow \tilde{H} h$ followed by $\tilde{H} \rightarrow h \tilde{G}$. In the top right panel the higgsino is heavier than the axino so the relevant decay chain is $\tilde{a} \rightarrow s \tilde{G}$ followed by $s \rightarrow hh/ZZ/WW$.  In the lower two panels the axino is the NLSP and the only decay mode is $\tilde{a} \rightarrow a \tilde{G}$, which is harmless.  Note that, in both purple and magenta regions, the exclusion from BBN is coming from the long lifetime of the decay to gravitinos at large $m_{3/2}$. 

In the lower two panels, the axino mass is sufficiently small that a new saxion decay channel opens up: $s \rightarrow \tilde{a} \tilde{G}$.  This leads to excessive dark matter in the large orange regions, again showing the critical importance of the axino mass in DFSZ theories.  Note there is a complementarity between $s \rightarrow \tilde{a} \tilde{G}$ (for $m_{\tilde{a}} < m_s$) leading to excessive dark matter and $\tilde{a} \rightarrow s \tilde{G}$ (for $m_s > m_{\tilde{a}} $) leading to BBN problems.

We have not displayed a panel for the spectrum $m_s>m_{\tilde{a}} >\mu$, where saxions decay into axinos (giving an orange over-closure region), and axinos decay into higgsinos (giving a purple BBN region).
In this case, the excluded regions are roughly the unions of these regions from the upper and lower panels of Figure~\ref{fig:gravDFSZ}.
Since we require $m_s < 2 \mu$, this applies to a small range of $m_s$.

\subsection{A Limit on the Saxion Condensate from Axion Parametric Resonance}
\label{subsec:parametric resonance}

The constraints derived in the previous sub-section require only the two mild assumptions declared early in Sec.~\ref{sec:history}.  To discuss the constraint from axion parametric resonance we add an assumption about evolution well above the TeV scale:
\begin{itemize}
\item  The saxion condensate oscillates about the present vacuum in a quadratic potential from an initial amplitude larger than $V_{\rm PQ}$.
\end{itemize}
In this case the saxion oscillation, through its coupling with the axion field, enhances the fluctuation of the axion field via the parametric resonance effect~\cite{Kofman:1994rk}.
Once the fluctuation of the axion field becomes much larger than $V_{\rm PQ}$, the axion field takes spatially varying random values, leading to the formation of domain walls after the QCD phase transition~\cite{Tkachev:1995md,Kasuya:1996ns,Kasuya:1997ha,Tkachev:1998dc,Kawasaki:2013iha}.
If $N_{\rm DW}=1$, these domain walls are unstable and decay into axions.
If $N_{\rm DW}>1$, these domain walls are stable, dominate the energy density of the universe, and hence are excluded.

For illustration, we consider the ``Decoupled Saxion" of Sec.~\ref{subsec:DecpSax}.
Saxion oscillations significantly enhance the fluctuations of the axion field modes with physical wave numbers around $k \sim m_s$ via parametric resonance.
When the saxion field has dropped from $s_I$ to $s$ and undergone $N_{\rm osc}(s)$ oscillations, the fluctuation of the angular direction is given roughly by
\begin{align}
\label{eq:axion fluctuation}
\delta \theta (s) \sim \left( \frac{H_{\rm inf}}{2\pi s_I} \right) e^{\mu N_{\rm osc}(s) },
\end{align}
where $e^\mu$ ($\mu = \mathcal{O}(1)$) is the growth rate per oscillation.
Here the factor of $(H_{\rm inf}/2\pi s_I)$ is the primordial fluctuation of the angular direction produced during inflation. Assuming the universe is radiation-dominated during these early oscillations,
\begin{align}
\label{eq:Nosc}
N_{\rm osc}(s) \sim \frac{m_s}{H(s)} = \left( \frac{s_I}{s} \right)^{ \scalebox{1.01}{$\frac{4}{3}$} }.
\end{align}
The number of the oscillations grows at small $s$, as the Hubble scale is smaller.

However, this scaling breaks down for $s \lsim V_{\rm PQ}$.  Additional fluctuations created once $s$ falls below $V_{\rm PQ}$ are small, $\Delta \theta < \mathcal{O}(1)$.
This can be seen easily by energy conservation. The energy density of the fluctuation is
\begin{align}
\rho_{\delta \theta} \sim  k^2 \Delta \theta^2 V_{\rm PQ}^2   \sim  m_s^2 \Delta \theta^2 V_{\rm PQ}^2.
\end{align}
and cannot be larger than the energy density of the saxion oscillations, $m_s^2 s^2$. Thus $\Delta \theta < \mathcal{O}(1)$ for fluctuations produced at $s<V_{\rm PQ}$.   The growth in the fluctuation cuts off as $s$ drops below $V_{\rm PQ}$, and the condition for domain walls not to be produced is  $\delta \theta (s \sim V_{\rm PQ})< \mathcal{O}(1)$, giving an upper bound on $s_I$,
\begin{align}
\label{eq:sI_upper bound}
\frac{s_I}{V_{\rm PQ}} < \left[ \frac{1}{\mu}{\rm ln} \frac{2\pi s_I}{H_{\rm inf}}  \right]^{ \scalebox{1.01}{$\frac{3}{4}$} } \sim \mathcal{O}(1\mathchar`-10).
\end{align}
Below the dashed lines in Figs.~\ref{fig:gravKSVZ} and~\ref{fig:gravDFSZ} with labels ``$s_I = 10 \, V_{\rm PQ}$", this condition is violated and we expect axion fluctuations with $\delta \theta > \mathcal{O}(1)$.
For $N_{\rm DW}>1$, the regions below the dashed lines are excluded.
For $N_{\rm DW}=1$, the axion misalignment angle is randomized  and these regions are subject to the constraint given in Figure~\ref{fig:axion}, which is shown by pink shadings.  These constraints may be avoided in cosmologies that violate the assumption itemized at the beginning of this sub-section. 

In the DFSZ theory with $N_{\rm DW}>1$, from Figure~\ref{fig:gravDFSZ} the only allowed parameter region has large $V_{\rm PQ}$.
In four dimensional grand unified theories, symmetries which control the $\mu$ term of the Higgs doublets must be broken at the unification scale~\cite{Goodman:1985bw,Witten:2001bf,Harigaya:2015zea}.
It is illuminating that the constraint from parametric resonance also points towards a large PQ breaking scale.

Note that this constraint is derived by evaluating (\ref{eq:axion fluctuation}) and (\ref{eq:Nosc}) at $s \sim V_{\rm PQ}$; 
the details of the evolution prior to this is irrelevant.  Thus the constraint applies provided the saxion condensate oscillates about the present vacuum in a quadratic potential from an initial amplitude larger than $V_{\rm PQ}$.   However, to phrase the constraint in a way that is independent of the earlier saxion evolution requires a reinterpretation of $s_I$ in Eq.~(\ref{eq:sI_upper bound}).
 On the left hand side of Eq.~(\ref{eq:sI_upper bound}), $s_I$ is a parametrization of the strength of the saxion condensation, and can be rewritten in terms of $T_{\rm NA}$ via Eq.~(\ref{eq:TNA}). In this more general formulation of the bound, the positions of the dashed lines and pink shaded regions are not changed.
In the log in the middle of Eq.~(\ref{eq:sI_upper bound}), $s_I$ parametrizes the size of the primordial fluctuation of the angular direction, and should be replaced by the PQ symmetry breaking scale during inflation. It affects the bound only logarithmically.

\section{Signals}
\label{sec:signals}

\subsection{Displaced Vertices, Kinks and Saxion Resonances at Colliders}
\label{subsec:sigDV}

\begin{table}
     \begin{center}
     \begin{tabular}{ | c || c | c | }
     \hline
          & $\chi^0$ LOSP & $\tilde{\tau}_R$ LOSP \\ \hline
      DFSZ &
      \raisebox{-.4\height}{\includegraphics[width=0.22\textwidth]{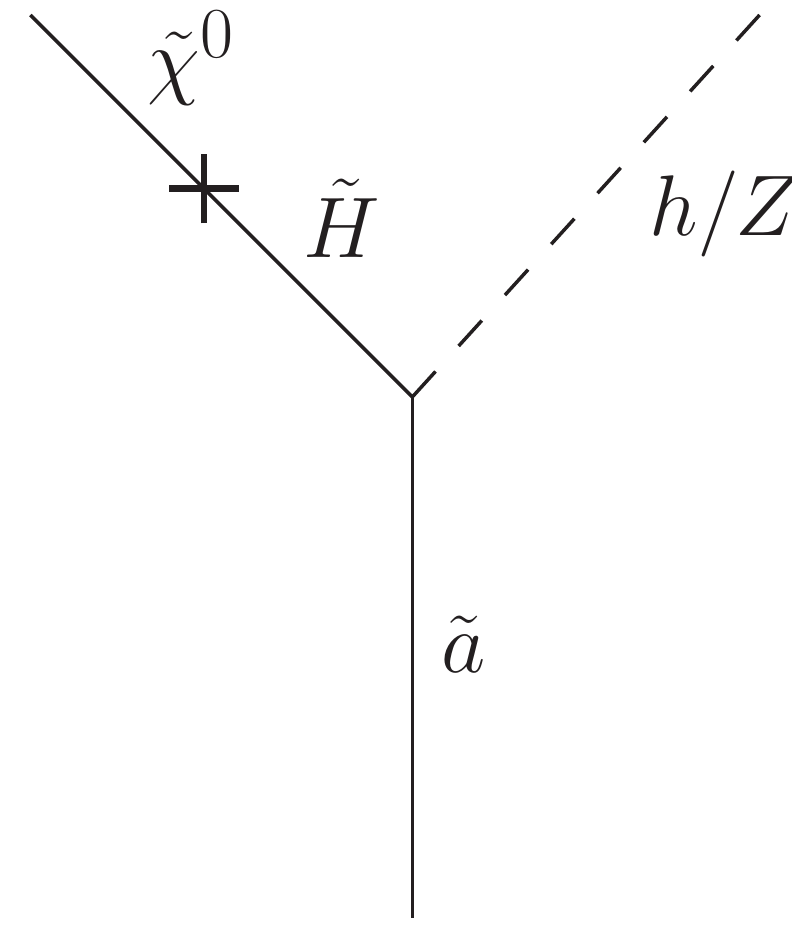}}
      & 
     \raisebox{-.4\height}{\includegraphics[width=0.22\textwidth]{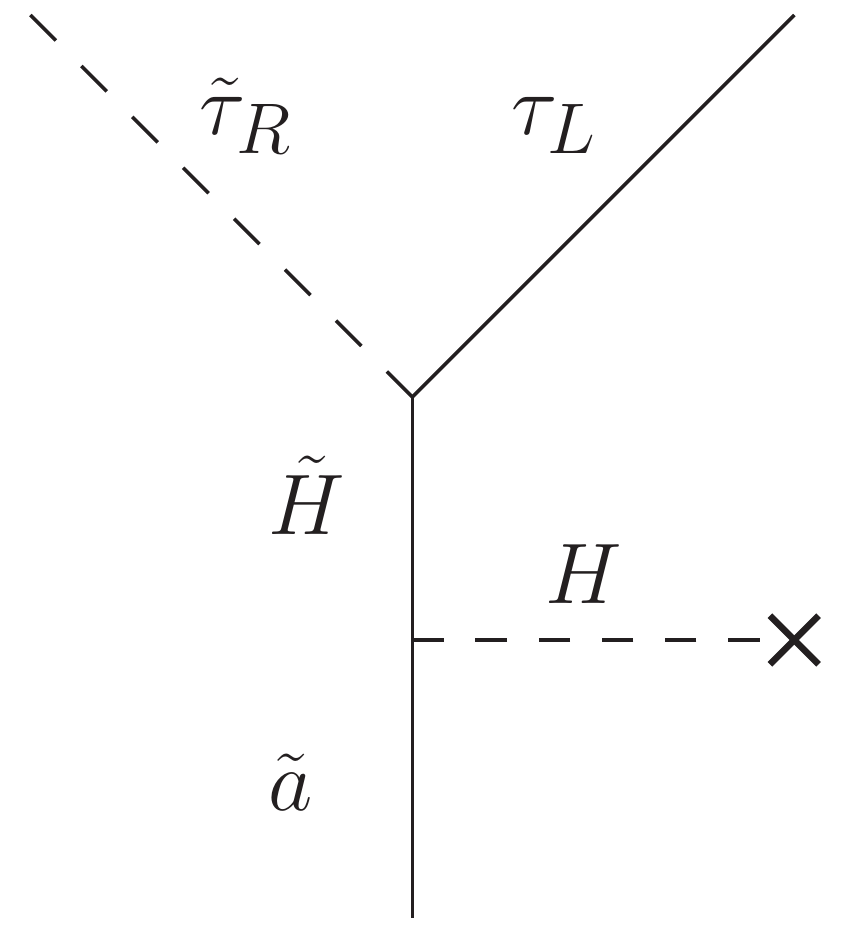}  \hspace{0.5in} \includegraphics[width=0.22\textwidth]{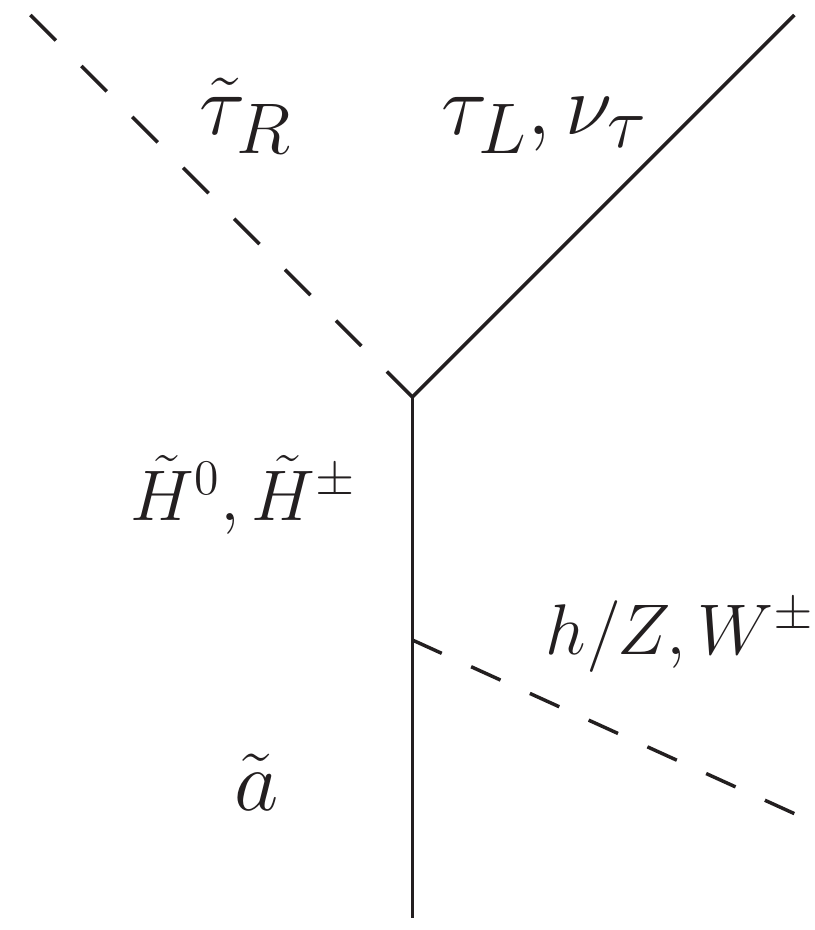}}
      \\ 
      & (a) & (b)  \hspace{4.4cm}  (c)
      \\ \hline 
      
      KSVZ &
      \raisebox{-.4\height}{\includegraphics[width=0.22\textwidth]{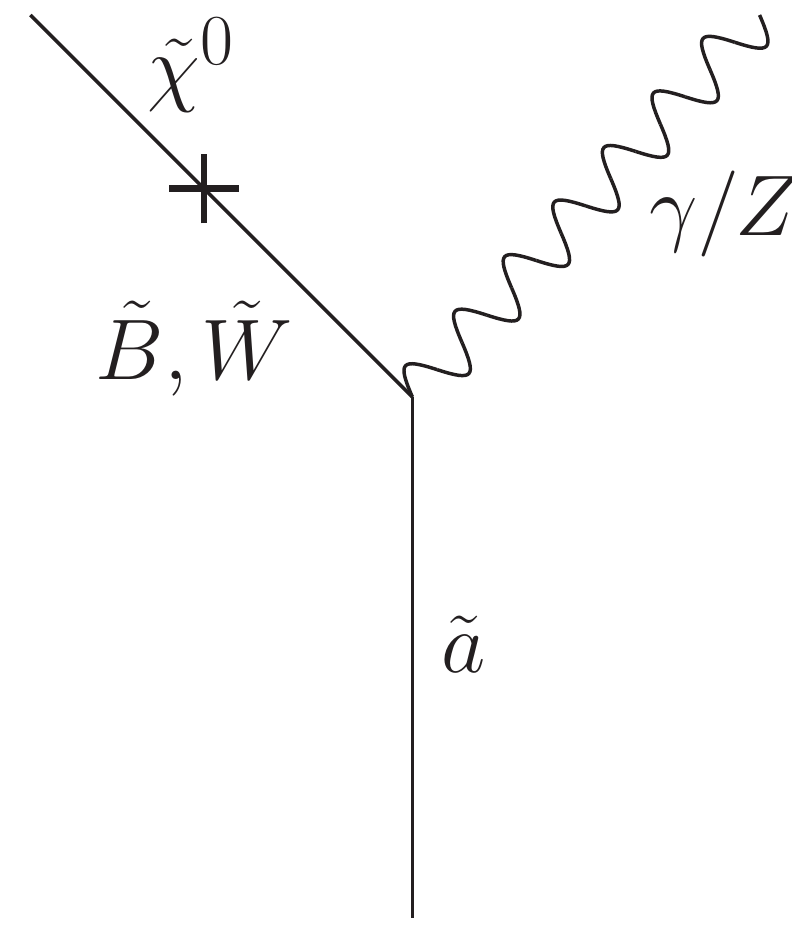}}
      & 
      \raisebox{-.4\height}{\includegraphics[width=0.22\textwidth]{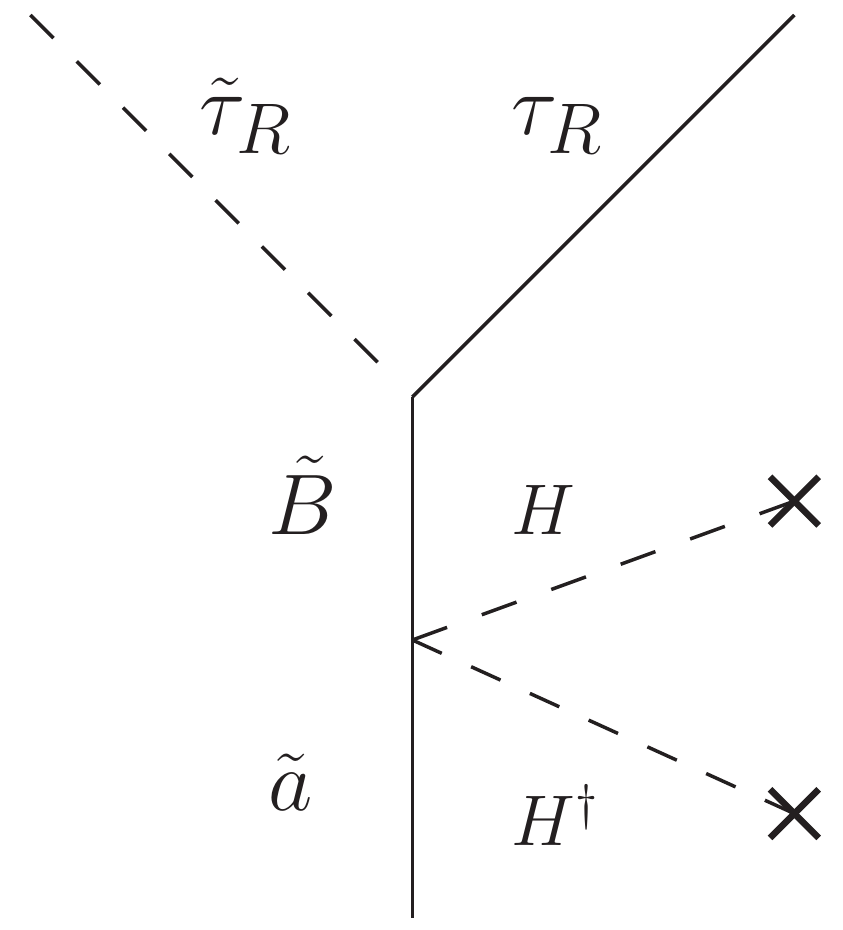}  \includegraphics[width=0.22\textwidth]{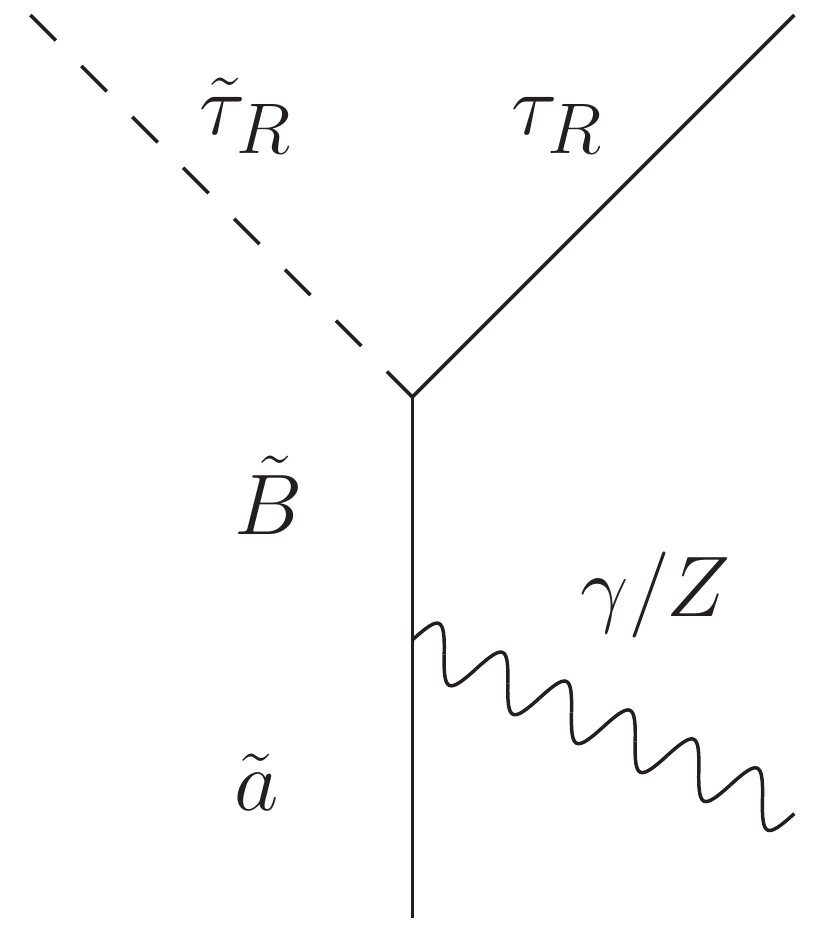} \includegraphics[width=0.22\textwidth]{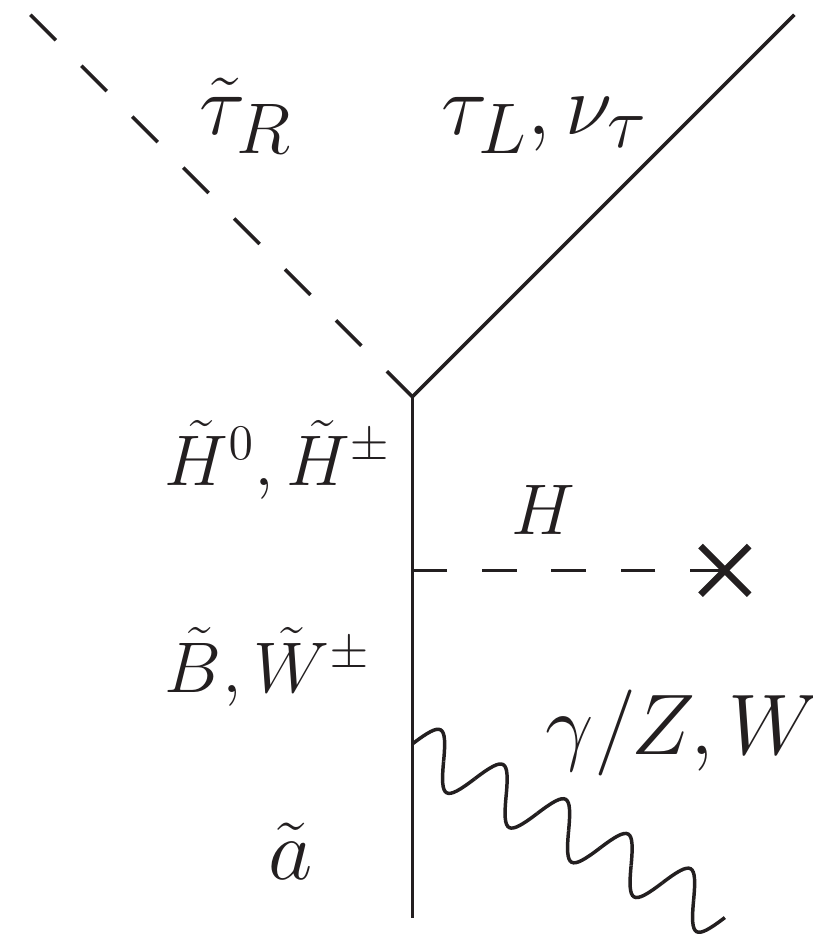}}
      \\ 
      \hspace{0.2cm}  & (d) & (e)  \hspace{3cm}  (f) \hspace{3.1cm}  (g)
      \\ \hline 
      \end{tabular}
      \caption{Displaced signals from $\tilde{\chi}^0$ and $\tilde{\tau}_R$ LOSPs decaying to NLSP axinos.}
      \label{table:DisplacedAxinos}
      \end{center}
      \end{table}

We discuss the following colliders signals resulting from LOSP decays -- displaced vertices and kinks involving gravitinos or axinos, and the saxion resonance. For illustration we assume a neutralino ($\tilde{\chi}^0$) or right-handed stau ($\tilde{\tau}_R$) LOSP.

The conventional signals of displaced vertices involving gravitinos result from NLSP decay via interactions suppressed by the mediation scale of supersymmetry breaking, with a decay length given by 
\begin{align}
c\tau_{\rm NLSP \rightarrow \tilde{G}} \equiv \frac{c}{\Gamma_{{\rm NLSP} \rightarrow \tilde{G}}} \simeq  2 \, \text{m} \; \left( \frac{ \TEV }{ m_{\rm NLSP} }  \right)^5  \left( \frac{ m_{3/2} }{ 100 \ \text{keV} }  \right)^2 .
\label{eq:NLSPctau}
\end{align}
Examples include the neutralino (stau) NLSP decaying to a gravitino and $h, \gamma, Z \; (\tau)$. This conventional signal applies whenever the LOSP decays to the axino are kinematically forbidden or sufficiently suppressed.   In Figure 2 and the upper left panel of Figure~\ref{fig:gravDFSZ} the axino is heavier than the LOSP, and in Figure~\ref{fig:gravKSVZ} the axino may be heavier than the LOSP.    Hence, in all these cases a displaced signal from (\ref{eq:NLSPctau}) can occur.  In the lower left panel of Figure~\ref{fig:gravDFSZ}, this signal competes with LOSP decays to axinos, described below.

We may also observe displaced vertices and/or kink signals if the axino is lighter than the LOSP. Any MSSM particles produced at the colliders will first cascade down to the LOSP. Through the axino interactions with higgsinos (gauginos) for DFSZ (KSVZ) theories, the LOSP will decay into the axino and SM particles.

In DFSZ models with a neutralino LOSP, the neutralino decays to the axino and the Higgs/$Z$ boson -- (a) of Table~\ref{table:DisplacedAxinos} -- with 
\begin{align}
c\tau_{\tilde{\chi}^0 \rightarrow \tilde{a}} \equiv \frac{c}{\Gamma_{\tilde{\chi}^0 \rightarrow \tilde{a}}} \simeq  5 \, \text{m} \; \left( \frac{2}{q_\mu} \right)^2 \left( \frac{ \mu }{ m_{\tilde{\chi}^0} }  \right)  \left( \frac{ 10^3 \ \text{GeV} }{\mu}  \right)^3  \left( \frac{ V_{\rm PQ} }{10^{12} \ \text{GeV}}  \right)^2  C_{\tilde{\chi}^0\tilde{H}}^{-2} \ ,
\label{eq:Higgsinoctau}
\end{align}
where $C_{\tilde{\chi}^0\tilde{H}}$ is the higgsino component of the neutralino LOSP~\cite{Co:2015pka}. This is applicable to all panels of Figure~\ref{fig:gravDFSZ}, except the upper left since we need $\mu > m_{\tilde{a}}$. Below the gray regions, this decay mode is typically more efficient than that to the gravitino final state, as can be seen from the smaller suppression scale $m_s V_{\rm PQ} \ll m_{3/2} M_{\rm Pl}$. Thus the decay of the neutralino into the axino and the Higgs/$Z$ boson may be observed as a displaced vertex.
With the stau LOSP, $\tilde{\tau} \rightarrow \tau \tilde{a}$ through higgsino-axino mixing  -- (b) of Table~\ref{table:DisplacedAxinos} -- with
\begin{align}
c \tau_{\tilde{\tau} \rightarrow \tilde{a}} \simeq 1~{\rm m} \left( \frac{2}{q_\mu} \right)^2 \left(\frac{10^3 \text{GeV}}{m_{\tilde{\tau}}}\right)  \left( \frac{ V_{\rm PQ} }{10^{10} \ \text{GeV}}  \right)^2 \left( \frac{10}{{\rm tan}\beta} \right)^2.
\end{align}
Similar to conventional gauge mediation with a stau NLSP, this decay may leave a kink signal. For large stau masses, the stau instead dominantly decays to $\nu_\tau + W + \tilde{a}$ or $\tau + Z/h + \tilde{a}$ -- (c) of Table~\ref{table:DisplacedAxinos}. The latter decay is observed as a kink where $Z/h$ is emitted.

In KSVZ models, a neutralino LOSP decays to $(\gamma/Z) \, \tilde{a} $ -- (d) of Table~\ref{table:DisplacedAxinos} -- and is observed as a displaced vertex.
Since this mode is loop-induced the decay rate is smaller that the one in Eq.~(\ref{eq:Higgsinoctau}), typically by a factor of $10^{5\mathchar`-6}$.
On the other hand, a stau LOSP decays into $\tau_R \, \tilde{a}$ -- (e) of Table~\ref{table:DisplacedAxinos} -- through axino-bino mixing arising from the non-zero electroweak $D$ term, leaving a pure kink. 
For large stau masses this mixing becomes quadratically suppressed by the electroweak vev, so that the 3-body final state $\tau_R \, (\gamma/Z) \, \tilde{a} $ becomes favored -- (f) of Table~\ref{table:DisplacedAxinos}.
The stau also has 3-body decays linearly suppressed by the electroweak vev:  $\tau_L \, (Z/\gamma) \, \tilde{a}$ and $\nu_\tau \, W \, \tilde{a}$ -- (g) of Table~\ref{table:DisplacedAxinos} -- the latter has $W$ appearing at a displaced vertex. All of the above modes are suppressed by loop factors as well as three-body phase space factors or the ratio between the electroweak and SUSY scales. The decay of the LOSP into axinos is sub-dominant near the gray-shaded regions of Figures \ref{fig:gravMax}, \ref{fig:gravKSVZ} and \ref{fig:gravDFSZ} and is dominant far enough from these regions.

If the saxion is heavier than the axino then the axino decays invisibly to an axion and a gravitino.   
However, an interesting and unique signal arises when $m_s < m_{\tilde{a}} < m_{LOSP}$, e.g.~in the upper right panel of Figure~\ref{fig:gravDFSZ}.  Since the lower limit on the saxion mass is of order (10 MeV, 1 GeV) for (DFSZ, KSVZ) theories, this can occur in a wide region of parameter space.  In this case, the LOSP is produced at the collision point, travels a distance $c\tau_{\rm LOSP \rightarrow \tilde{a}}$ and decays to the axino and SM particles, leaving a displaced vertex or a kink if $c\tau_{\rm LOSP \rightarrow \tilde{a}}$ is in an appropriate range. The axino then travels some other distance $c\tau_{\rm NLSP \rightarrow \tilde{G}}$ before it finally decays to the gravitino and the saxion/axion, with the saxion decaying to $hh/WW/ZZ/ \bar{f} f/gg$ for DFSZ and $gg$ for KSVZ, leaving a displaced vertex for an appropriate $c\tau_{\rm NLSP \rightarrow \tilde{G}}$. Remarkably, the saxion can be observed as a resonance despite its feeble coupling with the SM. This particular decay mode has a distinctive feature of multi-jets from Higgs/$Z$ boson resonances, taus, missing energy, and a saxion resonance.

\subsection{A Warm Component of Dark Matter}
\label{subsec:sigWDM}

We have considered three sources for gravitinos: decoupling from the thermal bath, freeze-in by higgsino decays, and decays of freeze-in axinos. Given the observed DM abundance, dilution is always large enough that the thermally decoupled gravitinos satisfy warm DM constraints if $m_{3/2} \gsim \mathcal{O}(\KEV)$ \cite{Viel:2013apy}. At larger $m_{3/2}$ these gravitinos rapidly become cold.

On the other hand, gravitinos produced from the FI decays of higgsinos can be warm even if $m_{3/2}$ is larger than a keV.  These FI gravitinos give a significant component of DM only at low $V_{PQ}$ and $m_{3/2}$, for example near the boundary of the brown regions of Figures~\ref{fig:gravKSVZ} and \ref{fig:gravDFSZ}.  For this range of parameter space, FI gravitinos are produced in the MD$_{\rm NA}$ era and hence are diluted less than thermal gravitinos, leading to the larger free-streaming length of Eq.~(\ref{eq:FSFI}), as discussed in App.~\ref{app:WDM}. If FI gravitinos dominate DM, the warm DM bound on $m_{3/2}$ becomes somewhat more stringent, as shown in Eq.~(\ref{eq:mWDM}). For $m_{3/2} \sim \mathcal{O}(10 \, \KEV)$ near the brown regions of Figures~\ref{fig:gravKSVZ} and \ref{fig:gravDFSZ}, we predict a mixture of cold and warm dark matter from thermal and FI gravitinos, respectively.

In DFSZ theories, the abundance of gravitinos from FI axinos can be comparable to that of thermally decoupled gravitinos near the green regions of Figure~\ref{fig:gravDFSZ}. The FI axinos become non-relativistic before decaying to gravitinos, giving them momenta larger than the thermal gravitinos. When $m_{3/2} \ll m_{\tilde{a}}$, the free streaming length of gravitinos becomes independent of $m_{3/2}$ \cite{Co:2016fln} and can be approximated by
\begin{align}
\lambda_{\rm FS}^{\rm decay} \approx \, 1 \text{ Mpc } \left( \frac{650 \, \GEV}{m_{\rm NLSP}} \right)^{ \scalebox{1.01}{$\frac{3}{2}$} }  \left[1 + 0.15 \log\left(  \frac{m_{\rm NLSP}}{650 \, \GEV} \right)  \right] .
\end{align}
where the NLSP is the higgsino (axino) in the left (right) panels of Figure~\ref{fig:gravDFSZ}.
The analyses from Lyman-$\alpha$ forest \cite{Boyarsky:2008xj, Harada:2014lma, Kamada:2016vsc} place an upper bound on the dark matter free streaming length, $\lambda_{\rm FS} \lsim 1$~Mpc. As a result, for $m_{\rm NLSP} \approx 650$ GeV there is a sizable component of warm dark matter in the parameter space close to the boundaries of the green regions.

These warm gravitinos lead to possible signals in the Lyman-$\alpha$ observations. It has been shown that warm dark matter can solve the small scale structure problems although baryon feedback may also play a role. (See \cite{Weinberg:2013aya} for a review.)

\subsection{Dark Radiation}
\label{subsec:sigDR}

Axions may contribute to dark radiation in both KSVZ and DFSZ theories because the saxion can decay to a pair of relativistic axions via the trilinear coupling in the K\"ahler potential \cite{Co:2016vsi,Choi:1996vz}. This decay rate depends on the model-dependent parameter $\kappa = \sum_i q_i^3 v_i^2/V_{\rm PQ}^2$
\begin{align}
\Gamma_{s\rightarrow aa} = \frac{\kappa^2 \, m_s^3}{64 \pi V_{\rm PQ}^2},
\end{align}
where $q_i$ and $v_i$ are the PQ charge and vev of each PQ breaking field, leading to an effective number of relativistic neutrinos 
\begin{align}
\Delta N_{\rm eff} = \frac{4}{7} \, g_*(T_{\nu \, dec}) \frac{\Gamma_{s\rightarrow aa}}{\Gamma_{s}} = 
\begin{cases} 
      \, 0.5 \, \left( \frac{\kappa}{0.1} \right)^2 \left( \frac{0.1}{\alpha_3} \right)^2 \left( \frac{10}{N_{\rm DW}} \right)^2 & \text{KSVZ} \\
      \, 0.1 \, \kappa^2 \,   \left(\frac{2}{q_\mu} \right)^2  \left(\frac{m_s}{\mu} \right)^4   & \text{DFSZ } (m_s > 2 \, m_h) \\
      \, 0.3 \, \kappa^2 \,   \left(\frac{2}{q_\mu} \right)^2  \left(\frac{100~{\rm GeV}}{\mu} \right)^4 \frac{N_f}{3} \left(\frac{m_s/2}{m_f}\right)^2   & \text{DFSZ } (m_s < 2 \, m_W).      
   \end{cases}  
   \label{eq:DR}
\end{align}
In KSVZ theories, we take $\kappa \lsim \mathcal{O}(0.1)$ to be compatible with the current Planck constraint of $\Delta N_{eff} = 0.6$~\cite{Ade:2015xua}. A small $\kappa$ can arise from an approximate $Z_2$ symmetry or fine tuning. For DFSZ theories, the constraint on $\kappa$ is much relaxed because the saxion decay to the visible sector is more efficient;
in fact, for $m_s< 2\mu$, as required to forbid the decay of the saxion into a pair of higgsinos, the Planck constraint is satisfied even if $\kappa \sim \mathcal{O}(1)$.  In all three cases of  (\ref{eq:DR}), part of parameter space is accessible to the CMB-S4 experiment.

\section{Conclusions}
The main results of this paper are shown in Figures~\ref{fig:gravMax}, \ref{fig:gravKSVZ} and \ref{fig:gravDFSZ}. Gravitinos that were thermalized early in the universe and later diluted from the decay of a saxion condensate provide an excellent candidate for dark matter; they are subject to several important constraints, but these leave large allowed regions in the $(m_{3/2}, V_{\rm PQ})$ plane; as large as shown in Figure~\ref{fig:gravMax}.   Results are shown for a very wide range of $(m_{3/2}, V_{\rm PQ})$ and saxion mass $m_s$, and are independent of almost all UV model-dependence.

In KSVZ theories a large part of the $(m_{3/2}, V_{\rm PQ})$ parameter space is allowed, although there is a strong upper bound on $V_{\rm PQ}$ for the saxion condensate to decay before BBN, as shown in Figure~\ref{fig:gravKSVZ}.
For this upper bound on $V_{\rm PQ}$ to be larger than $10^9$ GeV, the saxion mass must be larger than $\mathcal{O}(1)$ GeV.

In DFSZ theories the efficient interaction between the axion and Higgs multiplets weakens the upper bound on $V_{\rm PQ}$, but puts strong constraints on $(m_{3/2}, V_{\rm PQ})$ from a variety of processes, as shown by shading in Figure \ref{fig:gravDFSZ}.  However, these additional constraints can be removed, for example by making the axino mass sufficiently large.
The reduced upper bound on $V_{\rm PQ}$ allows for a saxion mass as small as $\mathcal{O}(10)$ MeV, where both BBN and astrophysical bounds are included.

If the saxion begins its oscillation with a field value larger than $V_{\rm PQ}$, parametric resonance may induce large fluctuations of the axion field.  In theories with domain wall number $N_{\rm DW}>1$, these large fluctuations lead to the formation of disastrous stable domain walls. This is avoided for large $V_{\rm PQ}$, which typically has a larger region in DFSZ theories, although in KSVZ theories $N_{\rm DW} = 1$ is more easily obtained.

We have also estimated the axion abundance from the misalignment angle.
For a sufficiently small saxion decay rate, the axion abundance is also diluted.
For $V_{\rm PQ}>10^{12}$ GeV and $\theta_{\rm mis}$ order unity, some dilution is required, placing an upper bound on the decay rate,
as shown in Figure~\ref{fig:axion}.
In DFSZ theories the higgsino mass is bounded from above accordingly.

Here we summarize possible signals of our scenario:
\begin{itemize}
\item 
If the LOSP is lighter than the axino, it will decay to light gravitinos at displaced vertices.   We have provided a cosmology with high $T_R$ for this well-known signal, for example, of low-scale gauge mediation.
\item
If the LOSP is heavier than the axino, the axion multiplet participates in the decay chain of MSSM particles produced at colliders (Table~\ref{table:DisplacedAxinos}),
leaving displaced vertices and/or kinks.
If the saxion is lighter than the axino it is produced through axino decay and can be observed as a resonance.
\item
In some parameter regions gravitinos are dominantly produced via freeze-in processes.
Such gravitinos may behave as warm dark matter even if $m_{3/2}>$ keV.
\item
The saxion condensate also decays to relativistic axions, leading to a non-zero dark radiation abundance.
\end{itemize}

A combination of measurements, especially from displaced vertices or kinks at colliders, could constrain the theory and narrow the prediction for $V_{PQ}$, which may allow an independent probe from axion physics.

\appendix
\section{Warm Dark Matter from Freeze-In Gravitinos}
\label{app:WDM}

If dark matter is initially thermalized and decouples from the bath while relativistic, $m_{\rm DM} < T_{\rm dec}$, its energy at decoupling is of order the decoupling temperature $T_{\rm dec}$ . A sufficiently light dark matter particle will affect structure formation via its large velocity at matter-radiation equality, leading to a lower bound on $m_{\rm DM}$~\cite{Viel:2013apy}. Nonetheless, this bound is different when dark matter is produced from freeze-in decays during a MD era instead of thermal decoupling during a RD era. In this section, we investigates how the freeze-in scenario affects the constraint. 

In general, the DM abundance today is related to the initial one at production by 
\begin{align}
\label{eq:rhoD}
\frac{\rho_{f}}{s_f} = \frac{m_{\rm DM} n_{i}}{s_i D} = \frac{m_{\rm DM} Y_{\rm th} \epsilon}{D},
\end{align}
where $D$ is the dilution factor, $s$ the entropy density, and the yield is parametrized by $\epsilon$ in units of the thermal equilibrium value $Y_{\rm th}$ of relativistic Weyl fermions. For freeze-in (thermal decoupling), $\epsilon < 1$ ($\epsilon = 1$). We are concerned with the free-streaming length of dark matter so we study how dilution affects the momentum red-shift;
\begin{align}
\frac{p_f^3}{s_f} = \frac{p_i^3}{s_i D} \simeq  \frac{p_i^3 \rho_{\rm DM} }{s_i  s_f m_{\rm DM} Y_{\rm th}  \epsilon} ,
\end{align}
where $D$ is substituted using Eq.~(\ref{eq:rhoD}). This gives the momentum $p_f$ at any temperature 
\begin{align}
\label{eq:pfDM}
\frac{p_f }{m_{\rm DM}}= p_i \left( \frac{ \rho_{\rm DM} }{s_i m_{\rm DM}^4 Y_{\rm th}  \epsilon} \right)^{ \scalebox{1.01}{$\frac{1}{3}$} }.
\end{align}

For gravitinos that originate from higgsino decays at the freeze-in temperature $T_{\rm FI} = m_{\tilde{H}}/x_{\rm FI}$, with $x_{\rm FI} \sim 2-5$, $p_i \simeq m_{\tilde{H}}/2$, whereas $p_i \simeq T_{\rm dec}$ for those that decouple from the thermal bath at temperature $T_{\rm dec}$. Specifically, the ratio $p_i/s_i^{1/3}$ in Eq.~(\ref{eq:pfDM}) becomes a constant for each production mechanism and is larger by a factor of $x_{\rm FI}/2$ for freeze-in than thermal decoupling. This implies that the free-streaming length in the case of freeze-in is enhanced in comparison. The constraint from Lyman-$\alpha$ gives a lower bound on $p_f/m_{\rm DM}$ and thus a lower bound on $m_{\rm DM} \propto (x_{\rm FI}/2)^{3/4} \epsilon^{-1/4}$. Although the freeze-in gravitino phase space distribution is different from that of thermally decoupled gravitinos, we expect a bound of similar order applies.  Therefore, the constraint on $m_{3/2}$ for freeze-in can be obtained from rescaling the result of Ref.~\cite{Viel:2013apy} that assumes thermal decoupling
\begin{align}
   \label{eq:mWDM}
m_{3/2} >
\begin{cases} 
      \, \mathcal{O}(\KEV) \   & \text{thermal decoupling} \\
      \, \left( \frac{x_{\rm FI}^3}{8 \, \epsilon} \right)^{ \scalebox{1.01}{$\frac{1}{4}$} } \mathcal{O}(\KEV) \  & \text{freeze-in.}  
   \end{cases}  
\end{align}
The free-streaming length can be computed using Eq.~(\ref{eq:pfDM})
\begin{align}
\label{eq:FSFI}
\lambda_{\rm FS}^{\rm FI} \sim \, 0.4 \text{ Mpc } \epsilon^{1/3} \left( \frac{\KEV}{m_{3/2}} \right)^{ \scalebox{1.01}{$\frac{4}{3}$} } .
\end{align}
One can estimate the yield for freeze-in production by
\begin{align}
\epsilon \simeq \frac{Y_{\rm FI}}{Y_{\rm th}} = \frac{\Gamma_{\tilde{H}\rightarrow \tilde{G}} H_{\rm FI} Y_{\tilde{H},\,FI} }{Y_{\rm th}} \ .
\end{align}

In the brown regions of Figs.~\ref{fig:gravKSVZ} and~\ref{fig:gravDFSZ}, FI gravitinos are overproduced. We investigate the parameter space immediately above these brown regions so that the warm FI gravitino abundance is sizable. In this case, gravitinos freeze-in during the MD$_{\rm A}$ era so 
\begin{align}
H_{\rm FI} \approx \frac{\pi \sqrt{g_*(T_{\rm FI})}}{3\sqrt{10}} \frac{\sqrt{T_{\rm NA}^5T_{\rm FI}^3}}{T_{{\rm R}s}^2 M_{\rm Pl}} \ ,
\end{align}
which we use to derive
\begin{align}
   \label{eq:epsilon}
\epsilon = \frac{15 \mu^{7/2}}{\pi^4m_{3/2}^2T_{\rm NA}^{5/2}} \frac{ Y_{\tilde{H},\,FI} }{Y_{\rm th}} \times
\begin{cases} 
         \, \frac{\alpha_3^2m_s^3}{32 \pi^2 f_a^2}  & \text{KSVZ} \\
	\, \frac{\mu^4}{m_s V_{\rm PQ}^2}   & \text{DFSZ } (m_s > 2 \, m_h) .
   \end{cases}  
\end{align}
Note that $\epsilon$ is necessarily less than unity because the FI yield cannot exceed the thermal value. If the above expression gives $\epsilon > 1$, gravitinos stay in thermal equilibrium until $T \sim m_{\tilde{H}}$, where they decouple as the higgsino LOSPs becomes non-relativistic and exponentially depleted. Using the numerical results of $T_{\rm NA}$ in Figs.~\ref{fig:gravKSVZ} and~\ref{fig:gravDFSZ}, one finds that $m_{3/2} \sim \mathcal{O}(10 \, \KEV)$ can lead to warm freeze-in gravitinos in the parameter space near the edges of the brown regions.

\section*{Acknowledgement}
We thank Masahiro Kawasaki and Satoshi Shirai for useful discussion.
This work was supported in part by the Director, Office of Science, Office of High Energy and Nuclear Physics, of the US Department of Energy under Contract DE-AC02-05CH11231 and by the National Science Foundation under grants PHY-1316783 and PHY-1521446.
R.C.~was supported in part by the National Science Foundation Graduate Research Fellowship under Grant No.~DGE 1106400.
F.D.~is supported by the U.S. Department of Energy grant number DE-SC0010107. L.H. is supported by the Simons Foundation.

\end{document}